\documentclass[printer]{gLMA2e}

\usepackage{graphicx,rotating}
\usepackage{amsmath,amssymb}
\usepackage{german}
\newcommand{\diag}{{\rm diag\,}}
\newcommand{\e}{{\rm e}}
\newcommand{\ri}{{\rm i}}

\newcommand{\tr}{{\rm tr\,}}

\newcommand{\C}{\mathbb{C}}

\newcommand{\T}{\textstyle}
\newcommand{\N}{\mathbb{N}}
\newcommand{\Z}{\mathbb{Z}}
\newcommand{\Q}{\mathbb{Q}}
\newcommand{\R}{\mathbb{R}}

\newtheorem{conjecture}{Conjecture}
\newtheorem{problem}{Problem}

\begin{document}
\doi{10.1080/03081080xxxxxxxxxxxxx}
\issn{1563-5139}
\issnp{0308-1087} \jvol{00} \jnum{00} \jyear{2007} \jmonth{0000}

\markboth{Dirr, Helmke, Kleinsteuber, Schulte-Herbr{\"u}ggen}
{Relative $C$"~numerical ranges}

\title{Relative $C$"~Numerical Ranges for Applications\\
in Quantum Control and Quantum Information}

\author{G. Dirr\thanks{Corresponding author: 
\tiny{ dirr@mathematik.uni-wuerzburg.de,
http://www2.mathematik.uni-wuerzburg.de}
},
U. Helmke, M. Kleinsteuber\\
Department of Mathematics, University of W{\"u}rzburg\\
97074  W{\"u}rzburg, Germany\\[4mm]
Th.  Schulte-Herbr{\"u}ggen\\
Department of Chemistry, TU Munich\\
85747 Garching, Germany}
 
\received{\today}

\maketitle


\begin{abstract}
Motivated by applications in quantum information and quantum control,
a new type of $C$"~numerical range, the relative $C$"~numerical range
denoted $W_K(C,A)$, is introduced. It arises upon replacing
the unitary group $U(N)$ in the definition of the classical $C$"~numerical
range by any of its compact and connected subgroups $K \subset U(N)$.

The geometric properties of the relative $C$"~numerical range
are analysed in detail. Counterexamples prove its geometry is more intricate
than in the classical case: e.g.  $W_K(C,A)$ is neither
star-shaped nor simply-connected. Yet, a well-known result on the
rotational symmetry of the classical $C$"~numerical range
extends to $W_K(C,A)$, as shown by a new approach based on Lie theory. 
Furthermore, we concentrate on the subgroup
$SU_{\rm loc}(2^n) := SU(2)\otimes \cdots \otimes SU(2)$,
i.e. the $n$-fold tensor product of $SU(2)$, which is of particular
interest in applications. In this case, sufficient conditions are derived
for $W_{K}(C,A)$ being a circular disc centered at origin of the complex
plane.  Finally, the previous results are illustrated in detail for
$SU(2) \otimes SU(2)$. 
\end{abstract}


\section{Introduction}
\label{intro}

$C$"~numerical ranges and $C$"~numerical radii \cite{gold:77a,li:94,ando:94}
naturally relate to optimization problems on the unitary orbit of
a given matrix. A particularly interesting class of applications
concerns quantum systems as shown in more detail in the accompanying
paper \cite{schuldihegla:07}. This is because quantum dynamics and
quantum control of closed Hamiltonian systems are governed by
the unitary group of the underlying Hilbert space and its subgroups.

For instance, in quantum information and quantum computation, where
spin-$\frac{1}{2}$ two-level systems are exploited as elementary
units---called quantum bits or qubits for short---the full dynamics
of $n$ qubits take place on
the unitary group $SU(2^n)$. This includes two types of
coherent time evolutions, (i) {\em within} each individual qubit independently
as well as (ii) {\em between} coupled qubits.
The former is brought about by so-called \emph{local actions}. Hence,
we refer to the corresponding subgroup as the subgroup of local time
evolution $SU_{\rm loc}(2^n)$. It takes the form of an $n$-fold tensor
product of $SU(2)$ with itself.
Clearly, matrices (states) that are themselves $n$-fold tensor-products
remain in this product form under conjugation with elements of
$SU_{\rm loc}(2^n)$.
Therefore, quantum optimization problems which can be expressed via 
local actions motivate in a most natural way the study of a subset of the
classical $C$"~numerical range, namely the {\em local $C$"~numerical range}.
It is but a special instance of the more general structure which we introduce
here as the {\em relative $C$"~numerical range}, where the full unitary group
is restricted to some arbitrary compact connected subgroup.

To be mathematically precise, let $U(N)$ denote the compact and connected
Lie group of all unitary matrices of size $N$,
i.e. $U \in U(N)$ if and only if  $U \in \C^{N \times N}$ with
$UU^{\dagger} = U^{\dagger}U = I_N$, where
$U^{\dagger}$ stands for the conjugate transpose. Moreover, let $SU(N)$ be
the subgroup of all special unitary matrices, i.e. 
$U \in SU(N)$ if and only if $U \in U(N)$ with $\det U = 1$. Finally, let
\begin{equation*}
SU_{\rm loc}(2^n):=
\underbrace{SU(2)\otimes \cdots \otimes SU(2)}_{\mbox{$n$-times}}
\end{equation*}
be the $n$-fold tensor product of $SU(2)$, which consists of all
$n$-fold Kronecker products of the from
\begin{equation*}
U_1\otimes \cdots \otimes U_n, \quad U_k \in SU(2)
\end{equation*}
for all $k = 1, \dots, n$. Following terminology from 
quantum information---as mentioned above---we refer to $SU_{\rm loc}(2^n)$ as the subgroup
of {\it local} unitary transformations.
Now, for arbitrary complex matrices $C, A \in \C^{2^n \times 2^n}$ the
{\em local $C$"~numerical range} of $A$ is defined as
\begin{equation}
\label{WCADef}%
W_{\rm loc}(C,A):=\big\{\tr(C^\dagger UAU^\dagger)\;\big|\;
U\in SU_{\rm loc}(2^n)\big\} \subset \C.
\end{equation}%

Obviously, $W_{\rm loc}(C,A)$ is a subset of the classical $C$"~numerical
range of $A$, denoted $W(C,A)$. However, its geometry is significantly
more intricate than in the classical case and has yet not been studied
in any systematic way, cf. \cite{ando:94}.
Analysing its basic properties leads naturally to the following slightly
more general object
\begin{equation}
\label{def:relnumrange}%
W_{K}(C,A):=\big\{\tr(C^\dagger UAU^\dagger)\;\big|\;
U\in K\big\},
\end{equation}%
which we call the \emph{relative $C$"~numerical range} of $A$. Here, $K$
may be any compact and connected subgroup of $U(N)$. 

Clearly, $W_K(C,A)$ is invariant under unitary similarities of $A$ and $C$
by elements of $K$. Moreover, with $W_K(C,A)$ being a continuous image of
$K$, it is compact and connected.
However, in contrast to the classical $C$"~numerical range, cf.
\cite{cheung:96a},  counterexamples show that $W_K(C,A)$ is neither
star-shaped nor simply connected. In general, the actual size and
shape of $W_K(C,A)$ are mostly unknown. 

The purpose of the present paper is to initiate the study
of the relative $C$"~numerical range and in particular its geometry.
Starting points for our investigations are well-known results on the
classical $C$"~numerical range. Therefore, we have summarized the
fundamental geometric properties of $W(C,A)$ at the beginning of
Section \ref{sec:2}. Subsequently,
the above mentioned counterexamples and straightforward consequences of
(\ref{def:relnumrange}) are collected in Subsection \ref{subsec:2.1}. 
The main result, a generalisation on rotationally symmetric $C$-umerical
ranges, is presented in Subsection \ref{subsec:2.2}. Particularly, the
original work by Li and Tsing \cite{li:91} is recovered as a
immediate corollary. In Section \ref{sec:3} we return to local
$C$"~numerical ranges. Here, the central conclusion is that rotationally
symmetric local $C$"~numerical ranges are essentially circular discs.
Finally, we illustrate the previous results by determining all complex
$(4 \!\times\!4)$-matrices, the local $C$"~numerical range of which is a
circular disc centered at the origin of the complex plane. These
computations lead to a conjecture on
unitray similarity to block-shift form of such matrices. 
Some concluding remarks on numerical issuse and perspectives for
future work can be found in Section \ref{sec:4}.


\section{The Relative $C$"~Numerical Range}
\label{sec:2}

In Section \ref{intro}, we introduced the $C$"~numerical range of
$A \in \C^{N \times N}$ relative to a compact and connected subgroup
$K$ of $U(N)$ as follows
\begin{equation}\label{relCnumrange}
W_K( C,A) := 
\big\{\tr(C^{\dagger}UAU^{\dagger}) \;\big|\; U \in K\big\}.
\end{equation}
For simplicity, we call $W_K( C,A)$ \emph{the relative $C$"~numerical range}
whenever any misinterpretation is excluded. 

Below, we aim at a better understanding of the geometry of the relative
$C$"~numerical range. In particular, we focus on geometric properties of
the classical $C$"~numerical range which are preserved by passing to
the relative one. For this reason, we first recall some basic facts
about the classical $C$"~numerical range. 


\subsection*{The Classical $C$"~Numerical Range: A Short Survey}

Let $C \in \C^{N \times N}$ be any complex matrix. The classical numerical
and $C$"~numerical range of $A\in \C^{N \times N}$ are defined by
\begin{equation}
\label{numrange}
W(A)
:=\big\{\tr(x^{\dagger}Ax) \;\big|\; x \in \C^{N}, \|x\|=1 \big\}
\end{equation}
and
\begin{equation}
\label{Cnumrange}
W(C,A):=
\big\{\tr(C^{\dagger}UAU^{\dagger}) \;\big|\; U \in U(N)\big\},
\end{equation}
respectively. Since any unitary transformation $U$ can be factored in the
form $U = \e^{\ri \varphi}U'$ with $\det U' = 1$, the set $W(C,A)$ does
not change, if we restrict $U$ to $SU(N)$.
Moreover, if $C := xx^{\dagger}$ for some $x \in \C^{N}$ with $\|x\|_2=1$
we have
\begin{eqnarray}
\label{num-Cnum}
W(C,A)
& = &
\big\{\tr(xx^{\dagger}UAU^{\dagger}) \;\big|\; U \in U(n)\big\}
\nonumber\\
& = &
\big\{\tr((U^{\dagger}x)^{\dagger}AU^{\dagger}x) \;\big|\; U \in U(n)\big\}\\
& = & 
\big\{\tr(y^{\dagger}Ay) \;\big|\; y \in \C^{n}, \|y\|=1 \big\}
\; = \; W(A).
\nonumber
\end{eqnarray}
Hence $W(C,A)$ coincides with $W(A)$, if $C$ is of the above form.
Here, we used the fact, that $U(N)$ acts transitively on the unit
sphere of $\C^N$. This will be of interest later on, if one wants to
define a numerical range of $A$ relative to $K$, see also Section
\ref{sec:3}, Remark \ref{rem:3.1}.

Probably the most basic geometric properties of the numerical and
$C$"~numerical range are compactness and connectedness, both following
from the fact that $W(A)$ and $W(C,A)$ are continuous images of compact
and connected sets. The first deep result on the
geometry of the numerical range, its convexity, was obtained
independently by Hausdorff \cite{haus:19} and Toeplitz \cite{toep:18}.
For a proof in a modern textbook we refer to \cite{gusrao:97}.
Later, the result was extended by Westwick \cite{west:75} and Poon
\cite{poon:80} to $C$"~numerical ranges for one of the operators $A$ or $C$
being Hermitian. Yet, in general $C$"~numerical ranges are non-convex except
for $N=2$, and even for normal matrices convexity may fail for $N > 2$, cf.
\cite{west:75}. However, by a result of Cheung and Tsing \cite{cheung:96a},
they are always star-shaped with respect to the star-center 
$(\tr C^{\dagger})(\tr A)/N$. 

The above mentioned theorems describe features which are in common to
all $C$"~numerical ranges. Another type of results aims at characterising
$C$"~numerical ranges $W(C,A)$ of particularly simple form such as an interval
or a circular disc.  Here, we mention only two results in this direction.

\begin{theorem}[\cite{vneumann:37, bro:88}\hspace{2ex}]
\label{thm:hermitian_case}
If $C,A \in \C^{N \times N}$ are Hermitian, then $W(C,A)$ degenerates to a
compact interval $I=[a,b]$ on the real line with 
\begin{equation}
b = \sum_{j=1}^{N}\alpha_j \gamma_j
\quad\mbox{and}\quad
a = \sum_{j=1}^{N}\alpha_j \gamma_{n-j},
\end{equation} 
where $\alpha_1 \geq \dots \geq \alpha_n$ and
$\gamma_1 \geq \dots \geq \gamma_n$ are the eigenvalues of $A$ and $C$,
respectively. 
\end{theorem}

\begin{theorem}[\cite{li:91}\hspace{2ex}]
\label{thm:disc}
Let $A \in \C^{N \times N}$ and let
${\cal O}_u(A):= \{UAU^{\dagger}\;|\; U \in U(N)\}$ denote the
unitary orbit of $A$. The following statements are equivalent:
\begin{enumerate}
\item 
The unitary orbit of $A$ satisfies
$\e^{\ri \varphi} {\cal O}_u(A) = {\cal O}_u(A)$ for all $\varphi \in \R$.
\item
The $C$"~numerical range of $A$ satisfies $\e^{\ri \varphi} W(C,A) = W(C,A)$
for all $C \in \C^{N \times N}$ and all $\varphi \in \R$, i.e. $W(C,A)$ is
rotationally symmetric for all $C \in \C^{N \times N}$.
\item 
The $A$-numerical range of $A$ satisfies $\e^{\ri \varphi} W(A,A) = W(A,A)$
for all $\varphi \in \R$, i.e. $W(A,A)$ is rotationally symmetric.
\item 
The matrix $A$ is unitarily similar to a block matrix
$M = (M_{kl})_{1 \leq k,l \leq m}$ such that all $M_{kk}$ are square
blocks and $M_{kl} = 0$ if $l+1 \neq k$.
\item 
$\C^N$ can be factored into a direct sum of mutually orthogonal
subspaces $\C^N = S_1 \oplus \dots \oplus S_m$ such that
$A(S_k) \subset S_{k+1}$ for $k = 1, \dots, m-1$ and $A(S_m) = 0$.
\item 
The $C$"~numerical range $W(C,A)$ is a circular disc in the complex plane
centered at the origin for all $C \in \C^{N \times N}$.
\item
$W(A,A)$ is a circular disc in the complex plane centered at
the origin.
\end{enumerate}
\end{theorem}

Theorem \ref{thm:hermitian_case} can be traced back to von Neumann.
For a proof we refer to \cite{bro:88, helmke}. Theorem \ref{thm:disc}
is a resume of Theorem 2.1 and Corollary 2.2 in \cite{li:91}. More on
$C$"~numerical ranges and the $C$"~numerical radii can be found in a
special issue of \emph{Linear and Multilinear Algbra} by Ando and Li
\cite{ando:94}.

\begin{remark}
\label{rem:2.1}
It is essential in Theorem \ref{thm:disc} to require the symmetry
condition in statement (b) and (f) \emph{for all} $C \in \C^{N \times N}$.
In fact, $W(C,A)$ can be a circular disc, although neither $A$ nor $C$
satisfy any of the above conditions, for instance, if $A$ is Hermitian
and $C$ skew-Hermitian. 
\end{remark}


\subsection{Basic Properties}
\label{subsec:2.1}

Our starting points for analysing the geometry of the relative $C$"~numerical
range are the above mentioned facts on $W(C,A)$. Obviously, $W_K(C,A)$ is
compact and connected by the same argument as before.
However, the previous results on convexity and star-shapedness fail as the
following three counterexamples show. In particular, the first and third
one prove that $W_K(C,A)$ is in general not even simply connected. 

Here, the first example exploits the fact that the subgroup $K$ itself
need not be simply connected, while the second and third one are based on the
fact that the product of convex or star-shaped subsets of $\C$ is in general
neither convex nor star-shaped.

\begin{example}
\label{ex1}
Let
\begin{equation}
K:=
\left\{
\begin{bmatrix}
\e^{\ri \varphi} & 0\\ 0 & \e^{-\ri \varphi}
\end{bmatrix}
\;\Bigg|\;
\varphi \in \R
\right\} \cong U(1)
\quad\mbox{and}\quad
A := C :=
\begin{bmatrix}
0 & 0\\ 1 & 0
\end{bmatrix}.
\end{equation}
Then we have
\begin{equation}
W_K(C,A) = \{\e^{\ri 2\varphi} \;|\;\varphi \in \R\}
\cong S^1.
\end{equation}
Obviously, $W_K(C,A)$ is not simply connected and therefore neither
star-shaped nor convex. However, this is not really surprising,
since $K\cong U(1)$ is a closed one-parameter subgroup and hence
itself not simply connected. 


\end{example}

\begin{example}
\label{ex2}
Let $K:=U(2) \otimes U(2)$ and let
\begin{eqnarray*}
A & := & A_1 \otimes A_2
\quad\mbox{with}\quad
A_1 :=
\begin{bmatrix}
1 & 0\\ 0 & -1
\end{bmatrix},
\quad
A_2 :=
\begin{bmatrix}
1+\ri & 0\\ 0 & 1-\ri
\end{bmatrix},\\\\
C & := & C_1 \otimes C_2
\quad\mbox{with}\quad
C_1 := C_2 := 
\begin{bmatrix}
1 & 0\\ 0 & 0
\end{bmatrix}.
\end{eqnarray*}
By the trace identity
$
\tr (A \otimes B) = \tr A \cdot \tr B$
for all $A \in \C^{N \times N}$ and $B \in \C^{N' \times N'}$
it is easy to see that 
$W_K(C,A)  =  W(C_1,A_1) \cdot W(C_2,A_2)$.
Moreover, as $A_j$ is normal and $C_j$ is Hermitian for $j=1,2$, a
straightforward calculation yields
\begin{eqnarray*}
W(C_1,A_1) =  [-1,1]
& \quad\mbox{and}\quad &
W(C_2,A_2) =  \{1 + \ri b \;|\; -1 \leq b \leq 1\}.
\end{eqnarray*}
Hence we obtain 
$$
W_K(C,A) = \{a + \ri b\;|\; -1 \leq a \leq 1,\; |b| \leq |a| \}
$$
which is obviously not convex. 


\begin{figure}[!Ht]
\begin{center}
\includegraphics[width=0.45\textwidth]{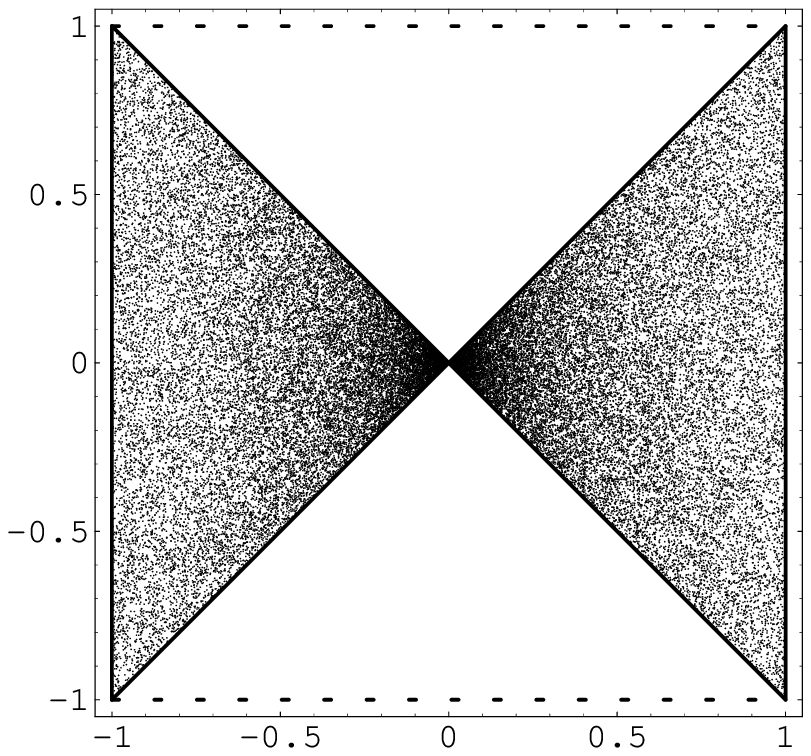}\quad
\raisebox{-.5mm}{
\includegraphics[width=0.4\textwidth]{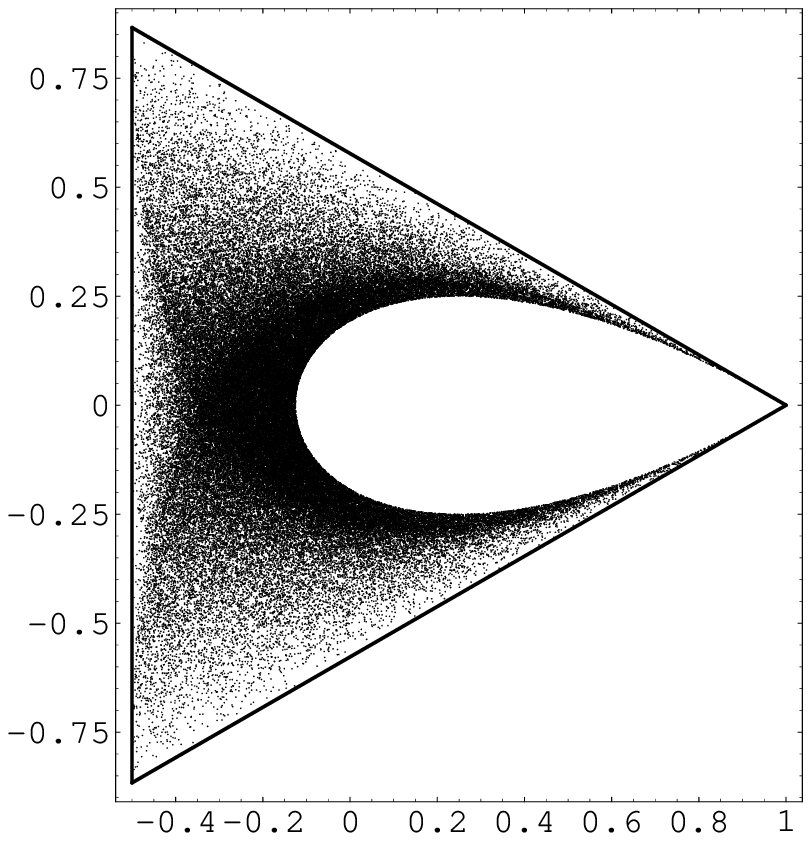}
}
\end{center}
\caption{\label{fig:bild23}
The relative $C$"~numerical ranges of Examples \ref{ex2} and \ref{ex3}.}
\end{figure}

\end{example}

\begin{example}\label{ex3}
Let $K:=U(2) \otimes U(2) \otimes U(2)$ and let
\begin{eqnarray*}
A & := & A_1 \otimes A_2 \otimes A_3
\quad\mbox{with}\quad
A_1 :=  A_2 := A_3 :=
\begin{bmatrix}
1 & 0\\ 0 & \e^{\frac{2\pi\ri}{3}}
\end{bmatrix},\\\\
C & := & C_1 \otimes C_2 \otimes C_3
\quad\mbox{with}\quad
C_1 := C_2 := C_3 :=
\begin{bmatrix}
1 & 0\\ 0 & 0
\end{bmatrix}.
\end{eqnarray*}
As in Example \ref{ex2}, we have 
$W_K(C,A) = W(C_1,A_1) \cdot W(C_2,A_2)\cdot  W(C_3,A_3)$
and
\begin{equation*}
W(C_j,A_j)  = 
\{(1-t) + t \e^{\frac{2\pi\ri}{3}} \;|\; t \in [0,1]\}
\quad\mbox{for}\quad j=1,2,3.
\end{equation*}
Thus, it is easy to see that the equilateral triangle 
$$
\Delta := W(C_1,A_1) \cup \e^{\frac{2\pi\ri}{3}}W(C_1,A_1)
\cup \e^{\frac{4\pi\ri}{3}}W(C_1,A_1)
$$ 
forms the outer boundary of $W_K(C,A)$, while the origin and a large
part of the interior of $\Delta$ does not belong to $W_K(C,A)$.
Therefore $W_K(C,A)$ is not simply connected and hence not
star-shaped either.


\end{example}

In both figures, the outer boundary of $W_K(C,A)$ has been computed
analytically and drawn in for a better visualization. In some sense,
Example \ref{ex3} is more significant than \ref{ex2}. Nevertheless,
we presented Example \ref{ex2}, because it seems to be impossible to
imitate \ref{ex3} in dimension $N=4$. This raises the question
which is the lowest dimension $N$ for which a compact, simply connected
subgroup $K \subset U(N)$ and matrices $A$, $C$ exist such that
$W_K(C,A)$ is not simply connected. Moreover, the reader should note
that both 
can be interpreted as relative numerical ranges, cf. Remark
\ref{rem:3.1} of Section \ref{sec:3}.

\bigskip

Having seen
that most of the geometric properties of the classical
$C$"~numerical range are not preserved, we present a series of results
which guarantee that some geometric features of $W(C,A)$ are passed
to $W_K(C,A)$ under additional assumptions on $A$, $C$, and
$K$. The first one is a trival fact based on the properties of the Hermitian
form $(A,C) \mapsto \tr C^{\dagger}A$.

\begin{lemma}
\label{HermitianLem}
Let $K$ be any compact and connected subgroup of $U(N)$ and let
$A,C$ be Hermitian. Then the $C$"~numerical range of $A$ is a compact
subinterval of $\R$. 
\end{lemma}

\begin{proof}
The identity $\overline{\tr C^{\dagger}A} = \tr A^{\dagger}C =
\tr CA^{\dagger} = \tr C^{\dagger}A $ for all Hermitian matrices $A,C$
implies immediately the above assertion. 
\end{proof}

The next two propositions focus on special types of subgroups---direct 
sums and tensor products---which are of particular interest for
applications, cf. \cite{schuldihegla:07}. Both results show that the
properties of the classical $C$"~numerical range are maintained, if the
matrices $A,C$ are compatible with the corresponding subgroup structure.

\begin{proposition}
\label{Prop:DirectSum}
Let $K_1 \subset U(N_1)$ and $K_2  \subset U(N_2)$ be compact and connected
subgroups and let $K$ be the direct sum of $K_1$ and $K_2$, i.e.
$K := K_1 \oplus K_1 := \{U_1 \oplus U_2 \;|\; U_1 \in K_1,\, U_2 \in K_2 \}$,
where $U_1 \oplus U_2 \in U(N_1+N_2)$ is defined by 
\begin{equation*}
U_1 \oplus U_2 :=
\begin{bmatrix}
U_1 & 0\\
0 & U_2
\end{bmatrix}.
\end{equation*}
Moreover, let $A := A_1 \oplus A_2$ and $C := C_1 \oplus C_2$ with
$A_i, C_i \in C^{N_i \times N_i}$ for $i = 1,2$. 
Then the $C$"~numerical range of $A$ is given by
\begin{equation}
\label{Eq:sum}
W_K(C,A) = W_{K_1}(C_1,A_1) + W_{K_2}(C_2,A_2).
\end{equation}
In particular, $W_K(C,A)$ is star-shaped (convex), if $W_{K_1}(C_1,A_1)$
and $W_{K_2}(C_2,A_2)$ are star-shaped (convex).
\end{proposition}

\begin{proof}
The result follows from 
$\tr (C_1 \oplus C_2)^{\dagger}(A_1 \oplus A_2) =
\tr C_1^{\dagger}A_1 + \tr C_2^{\dagger}A_2$
and the fact that the sum of two star-shaped (convex)
sets is again star-shaped (convex). 
\end{proof}

\begin{proposition}
\label{Prop:TensorProduct}
Let $K_1 \subset U(N_1)$ and $K_2  \subset U(N_2)$ be compact and connected
subgroups and let $K$ be the tensor product of $K_1$ and $K_2$, i.e.
$K := K_1 \otimes K_2 := 
\{U_1 \otimes U_2 \;|\; U_1 \in K_1,\, U_2 \in K_2 \}$,
where $U_1 \otimes U_2 \in U(N_1N_2)$ denotes the Kronecker product of
$U_1$ and $U_2$.
Moreover, let $A := A_1 \otimes A_2$ and $C := C_1 \otimes C_2$ with
$A_i, C_i \in C^{N_i \times N_i}$ for $i = 1,2$. 
Then the $C$"~numerical range of $A$ is given by
\begin{equation}
\label{eq:product}
W_K(C,A) = W_{K_1}(C_1,A_1) \cdot W_{K_2}(C_2,A_2).
\end{equation}
In particular, $W_K(C,A)$ is star-shaped (convex), if either
$W_{K_1}(C_1,A_1)$ is star-shaped (convex) and $W_{K_2}(C_1,A_2)$
is contained in a ray $\{r \e^{i \varphi} \;|\; r \geq 0\}$ of the
complex plane or vice versa. 
\end{proposition}

\begin{proof}
Eq. (\ref{eq:product}) is obtained by the trace identity
$\tr (A \otimes B) = \tr A \cdot \tr B$.
Moreover, $W_{K}(C,A)$ is star-shaped (convex),
if $W_{K_1}(C_1,A_1)$ is, as the following two operations preserve
star-shapedness and convexity: (a) rotation by a fixed complex number,
(b) multiplication by a subinterval of $\R^+_0$, cf. Appendix, Lemma
\ref{lem:starshaped}.
\end{proof}

\noindent
The following corollary is an immediate consequence of the previous
results.  

\begin{corollary}
Let $K$, $A$ and $C$ either be defined as in Proposition \ref{Prop:DirectSum}
or as in Proposition \ref{Prop:TensorProduct} and furthermore, let
$A_i, C_i$ be Hermitian for $i = 1,2$. Then
\begin{equation*}
W_K(C,A) = [a_1+a_2,b_1+b_2]
\vspace{-4mm}
\end{equation*}
or
\vspace{-4mm}
\begin{equation*}
W_K(C,A) = \big[\min\{a_1a_2,b_1b_2\},\max\{a_1a_2,b_1b_2\}\big],
\end{equation*}
respectively, where $W_{K_1}(C_1,A_1) = [a_1,b_1]$ and
$W_{K_2}(C_2,A_2) = [a_2,b_2]$.
\end{corollary}


\subsection{A Extension of the Circular-Disc Theorem}
\label{subsec:2.2}

This subsection contains the main result of the paper---a
generalisation of the Circular-Disc Theorem by Li and Tsing
\cite{li:91}, cf. Theorem \ref{thm:disc}. More precisely, we
characterize all matrices $A \in \C^{N \times N}$, the relative
$C$"~numerical range
of which is rotationally symmetric for all $C$.
But first, we illustrate by a counterexample that the {\em relative}
$C$"~numerical range of a block-shift matrix---as Li and Tsing's
result might suggest---is in general not rotationally symmetric.

\begin{example}\label{ex4}
Let $K:=U(2) \otimes U(2)$ and let
\begin{equation*}
A := C := 
\begin{bmatrix}
0 & 0 & 0 & 0\\
1 & 0 & 0 & 0\\
1 & 0 & 0 & 0\\
1 & 0 & 0 & 0\\
\end{bmatrix}.
\end{equation*}

\noindent
We claim that $W_K(C,A)$ is not circular, although its classical
$C$"~numerical range is a circular for all $C \in \C^{4 \times 4}$.
A rigorous proof of this fact will be given after Corollary
\ref{corlocrange}.
Moreover, we note that easier examples can be constructed via
the subgroup of Example \ref{ex1}. However, we favoured the
above one as it will be of interest also in Section \ref{sec:3}. 


\end{example}

Yet, for the main result and its proof, we need some further
notation. We call the $K$-orbit
$\mathcal{O}_K(A):= \{UAU^{\dagger} \;|\; U \in K\}$ of
$A \in \C^{N \times N}$ \emph{weakly rotationally symmetric}, if
\begin{equation*}
\e^{\ri \varphi} {\cal O}_K(A) = {\cal O}_K(A)
\end{equation*}
holds for all $\varphi \in \R$. Moreover, let $\mathfrak{k}$ denote
the Lie algebra of $K$ and let $\mathrm{ad}_{\Omega}$ for
$\Omega \in \mathfrak{k}$ be the operator defined by
\begin{equation*}
{\rm ad}_{\Omega}: \C^{N \times N} \to \C^{N \times N},
\quad A \mapsto {\rm ad}_{\Omega}(A) := [\Omega,A] := \Omega A - A \Omega.
\end{equation*}
Finally, a maximal Abelian subalgebra of $\mathfrak{k}$ is called a torus
algebra of $K$.

\begin{proposition}
\label{Prop:rotsym}
Let $A \in \C^{N \times N}$ and let $\varphi_0 \in \R$. Then the following
statements are equivalent:
\begin{enumerate}
\item
The orbit $\mathcal{O}_K(A)$ satisfies the relation
$\e^{i \varphi_0} \mathcal{O}_K(A) = \mathcal{O}_K(A)$.
\item
The relative $C$"~numerical range of $A$ satisfies
$\e^{\ri \varphi_0} W_K(C,A) = W_K(C,A)$ for all
$C \in \C^{N \times N}$.
\item 
The relative $A$-numerical range of $A$ satisfies
$\e^{\ri \varphi_0} W_K(A,A) = W_K(A,A)$.
\end{enumerate}
If $\varphi_0$ is an irrational multiple of $2\pi$ and if one of the above
statements (a), (b), or (c) holds for $\varphi_0$, then all of them are
satisfied for all $\varphi \in \R$.
\end{proposition} 

\begin{proof}
The implications (a) $\Longrightarrow$ (b) and (b) $\Longrightarrow$ (c)
are obvious. For proving (c) $\Longrightarrow$ (a) we assume without
loss of generality $A \neq 0$. Now, there is a $U_0 \in K$ such that
\begin{equation}\label{eq2.1}
\e^{\ri \varphi_0}\tr (A^{\dagger}A) = \tr (A^{\dagger}U_0AU_0^{\dagger}).
\end{equation}
Moreover, using the Cauchy-Schwarz inequality and the unitary invariance
of the Frobenius norm we obtain
\begin{equation}
\label{Eq:CauchySchwarz}
\|A\|^2 = |\e^{\ri \varphi_0}\tr (A^{\dagger}A)| = 
|\tr (A^{\dagger}U_0AU_0^{\dagger})| \leq
\|A\| \cdot \|U_0AU_0^{\dagger}\| = \|A\|^2.
\end{equation}
Thus, in fact equality holds in Eq. (\ref{Eq:CauchySchwarz}) and
therefore we have $U_0AU_0^{\dagger} = \lambda A$ for some
$\lambda \in \C$ with $|\lambda| = 1$. Substituting 
$U_0AU_0^{\dagger} = \lambda A$ in Eq. (\ref{eq2.1}) yields
$U_0AU_0^{\dagger} = \e^{\ri \varphi_0}A$ and hence
$\mathcal{O}_K(A) = \mathcal{O}_K(U_0AU_0^{\dagger}) =
\e^{\ri \varphi_0}\mathcal{O}_K(A)$. This finally implies (a).

\medskip
\noindent
If one of the above statements (a), (b) or (c) holds for some
$\varphi_0 \in \R \setminus 2\pi\Q$, we know that all of them
are satisfied for $\varphi_0$. In particular, by (a) we have 
\begin{equation}
\e^{\ri \varphi_0}A = U_0AU_0^{\dagger}
\end{equation}
for some $U_0 \in K$. Thus, by induction we obtain 
\begin{equation}\label{eq2.2}
\e^{\ri k \varphi_0}A = U_0^kA(U_0^k)^{\dagger}
\end{equation}
for all $k \in \Z$. Now, compactness of $K$ and Eq. (\ref{eq2.2}) 
imply that for all $\varphi \in \R$ there exists a $U \in K$
such that $\e^{\ri  \varphi}A = UAU^{\dagger}$
and thus 
\begin{equation}
\e^{\ri  \varphi}\mathcal{O}_K(A) = \mathcal{O}_K(A)
\end{equation}
for all $\varphi \in \R$.
Hence (a) and therefore also (b) and (c) are satisfied for all
$\varphi \in \R$.
\end{proof}

\begin{remark}
\label{Rem:2}
Proposition \ref{Prop:rotsym} is a slight generalization of the 
first part of Theorem 2.1 in \cite{li:91}. The above proof, which
is almost the same in \cite{li:91}, is included only for completeness.
\end{remark}

\noindent
As an immediate consequence of the above proposition we obtain.

\begin{corollary}
\label{Cor:rotsym}
Let $K \subset K'$ be compact and connected subgroups of $U(N)$.
\begin{enumerate}
\item 
The relative $C$"~numerical range of $A$ is rotationally symmetric
for all $C \in \C^{N \times N}$ if and only if the orbit $\mathcal{O}_K(A)$
is weakly rotationally symmetric.
\item 
If the relative $C$"~numerical range $W_K(C,A)$ is rotationally
symmetric for all $C \in \C^{N \times N}$, then the relative $C$"~numerical
range $W_{K'}(C,A)$ is as well. 
\end{enumerate}
\end{corollary}

\begin{proof}
(a) $\checkmark$

\medskip

\noindent
(b) This follows from (a) and the fact that weak rotational symmetry
of $\mathcal{O}_K(A)$ implies weak rotational symmetry of
$\mathcal{O}_{K'}(A)$.
\end{proof}

\noindent
Now, by the previous definitions our main theorem reads as follows.

\begin{theorem}
\label{Thm:rotsym}
Let $K$ be a compact and connected subgroup of $U(N)$ with Lie algebra
$\mathfrak{k}$ and let $\mathfrak{t}$ be a torus algebra of $\mathfrak{k}$.
Then the orbit $\mathcal{O}_K(A)$ of $A \in \C^{N \times N}$, $A \neq 0$
is weakly rotationally symmetric if and only if
\begin{enumerate}
\item[(a)]
there exists $\Omega \in \mathfrak{k}$
such that $A$ is an eigenvector of the operator ${\rm ad}_{\Omega}$
to a non-zero eigenvalue,
\end{enumerate}
or equivalently,
\begin{enumerate}
\item[(b)]
there exist $U \in K$  and $\Delta \in \mathfrak{t}$ such that
$UAU^{\dagger}$ is an eigenvector of the operator ${\rm ad}_{\Delta}$
to a non-zero eigenvalue.
\end{enumerate}
\end{theorem} 

Before proving Theorem \ref{Thm:rotsym}, the subsequent lemma will
clarify the relation between the operator ${\rm ad}_{\Omega}$ and 
the weak rotational symmetry of the orbit $\mathcal{O}_K(A)$.

\begin{lemma}\label{lemrotsym}
For $\Omega \in \mathfrak{u}(N)$ and $\varphi \in [0, 2\pi)$ let
$$
E_{\varphi}(\Omega):=
\{A \in \C^{N \times N} \;|\;
\e^{\ri\varphi t}A =
\e^{\Omega t}A\e^{-\Omega t}\;\mbox{\text{\rm for all} $t \in \R$}\}.
$$
Then $E_{\varphi}(\Omega)$ is equal to the eigenspace of the operator
${\rm ad}_{\Omega}$ to the eigenvalue $\ri \varphi$ , i.e. 
$$
E_{\varphi}(\Omega) =
\{A \in \C^{N \times N} \;|\;
{\rm ad}_{\Omega}(A) = \ri \varphi A\}.
$$
\end{lemma} 

\begin{proof}
``$\;\subset\;$'': Let
$A \in E_{\varphi}(\Omega)$. Then
\begin{equation}\label{eq2.3}
\e^{\ri\varphi t}A =
\e^{\Omega t}A\e^{-\Omega t}
\end{equation}
holds for all $t \in \R$. Hence differentiating Eq. (\ref{eq2.3}) on
both sides, yields 
$$
\ri \varphi A = \Omega A - A \Omega = {\rm ad}_{\Omega}(A)
$$ 
for $t=0$. Therefore, $A$ is an eigenvector of ${\rm ad}_{\Omega}$
to the eigenvalue $\ri \varphi$.

\medskip
\noindent 
``$\;\supset\;$'': Now, let $A$ be an eigenvector of
${\rm ad}_{\Omega}$ to the eigenvalue $\ri \varphi$. Then we
consider the curve $\omega(t) := \e^{\Omega t}A\e^{-\Omega t}$. An
easy calculation shows that $\omega(t)$ satisfies the linear
differential equation 
\begin{equation}\label{eq2.4}
\dot{\omega}(t) = \e^{\Omega t}(\Omega A - A \Omega)\e^{-\Omega t}
= \e^{\Omega t}{\rm ad}_{\Omega}(A)\e^{-\Omega t}
= \ri \varphi \omega(t).
\end{equation}
By uniqness of solutions of Eq. (\ref{eq2.4}) we have
$$
\omega(t) = e^{\ri \varphi} A \quad\mbox{for all $t \in \R$}
$$
and thus $A \in E_{\varphi}(\Omega)$.
\end{proof}

\begin{proof}[of Theorem \ref{Thm:rotsym}]
(a) ``$\;\Longleftarrow\;$'': Let $\Omega \in \mathfrak{k}$ and
assume that $A$ is an eigenvector
of the operator ${\rm ad}_{\Omega}$ to some eigenvalue
$\lambda \neq 0$. Then $\lambda$ is purely imaginary, i.e.
$\lambda = \ri \varphi$ for some $\varphi \in \R \setminus \{0\}$,
since ${\rm ad}_{\Omega}$ is skew-Hermitian with respect to the
scalar product $(A,C) \mapsto \tr (C^{\dagger}A)$. Therefore,
Lemma \ref{lemrotsym} implies
\begin{equation*}
\e^{\Omega t}A\e^{-\Omega t} = \e^{\ri\varphi t}A
\quad\mbox{for all $t \in \R$}
\end{equation*}
and this shows that $\mathcal{O}_K(A)$ is weakly rotationally
symmetric.

\medskip

\noindent ``$\;\Longrightarrow\;$'': Now, let $\mathcal{O}_K(A)$ be
weakly rotationally symmetric and choose $\varphi_0 \not\in 2\pi\Q$.
Thus there is a $U_0 \in K$ such that 
\begin{equation}\label{eq2.5}
\e^{\ri \varphi_0}A = U_0AU_0^{\dagger}
\end{equation}
and hence
\begin{equation}\label{eq2.6}
\e^{\ri k\varphi_0}A = (U_0^k)^{\dagger}AU_0^k
\quad \mbox{for all $k \in \Z$}.
\end{equation}
Now, we consider the closed Abelian subgroup $K_0$ generated by
$\{U_0^k \,|\, k \in \Z\}$. Since $K$ is compact, $K_0$ is the
direct product of a torus $T_0$ and a finite subgroup, c.f.
\cite{broedie}. Continuity and Eq. (\ref{eq2.6}) imply, that
for all $U \in T_0$ there exists a unique $\varphi$ depending on
$U$ such that
\begin{equation}
\label{eq2.7}
\e^{\ri \varphi}A = UAU^{\dagger}.
\end{equation}
Furthermore, let $\{\Omega_1, \dots, \Omega_m\}$ be a basis of the
Lie algebra of $T_0$ such that the following  conditions hold: 
\begin{equation}
\label{Eq:2.7a}
\e^{\Omega_j t} = I_N \;\mbox{for $t = 2\pi$}
\quad\mbox{and}\quad
\e^{\Omega_j t} \neq I_N \;\mbox{for $0 < t < 2\pi$}
\end{equation} 
for all $j = 1, \dots, m$. Since $U_0^k \in T_0$ for some $k \in \Z$,
we can assume $U_0 \in T_0$. Hence, there are $\alpha_j \in \R$ such that
$U_0 = \e^{\alpha_1\Omega_1 + \dots + \alpha_m\Omega_m}$.
Moreover, by Eq. (\ref{eq2.7}) there exist $\varphi_j \in \R$ for
$j = 1, \dots, m$ such that
\begin{equation}
\label{eq2.8}
\e^{\ri \varphi_j}A = \e^{\alpha_j\Omega_j}A\e^{-\alpha_j\Omega_j}.
\end{equation}
Now, from Eq. (\ref{eq2.5}) it follows that
at least one $\varphi_j$ is an irrational multiple of $2\pi$. Without
loss of generality let
$\varphi_1 \not\in 2\pi\Q$. Hence, we obtain
\begin{equation}
\e^{\ri \varphi_1}A = e^{\alpha_1\Omega_1}Ae^{-\alpha_1\Omega_1}
\end{equation}
and thus
\begin{equation}
\e^{\ri k \varphi_1}A = e^{\alpha_1\Omega_1 k}Ae^{-\alpha_1\Omega_1 k}
\end{equation}
for all $k \in \Z$. Finally, let $t \in \R$ and $(k_l)_{l \in \N}$ be a
sequence such that
$\lim_{l \to \infty}\e^{\alpha_1k_l\Omega_1} = e^{\Omega_1 t}$.
Since the map $\e^{\ri t} \mapsto \e^{\Omega_1 t}$
is a diffeomorphism of $S^1$ onto the one-parameter subgroup
$\{e^{\Omega_1 t} \;|\; t \in \R\}$ by Eq. (\ref{Eq:2.7a}), we conclude
$\lim_{l \to \infty}\e^{\ri \alpha_1k_l} = e^{\ri t}$ and so 
\begin{equation}
 e^{-\Omega_1 t}Ae^{\Omega_1 t}
= \lim_{l \to \infty}e^{-\alpha_1\Omega_1 k_l}Ae^{\alpha_1\Omega_1 k_l}
= \lim_{l \to \infty} \e^{\ri k_l \varphi_1}A =
\e^{\ri\frac{\varphi_1}{\alpha_1}t}A. 
\end{equation}
Therefore, Lemma \ref{lemrotsym} yields the desired result.

\bigskip

\noindent (b) ``$\;\Longrightarrow\;$'': If $UAU^{\dagger}$ is an
eigenvector of $\mathrm{ad}_{\Delta}$ to a non-zero eigenvalue
part (a) implies that the $K$-orbit of $UAU^{\dagger}$
is weakly rotationally symmetric. However, the $K$-orbit of $A$ and
$UAU^{\dagger}$ are equal. This proves the first part of (b). 

\medskip

\noindent ``$\;\Longleftarrow\;$'': Let $\mathcal{O}_K(A)$ be
weakly rotationally symmetric. Then by part (a) there exist
$\Omega \in \mathfrak{k}$ and $\varphi \in \R \setminus \{0\}$
such that
\begin{equation}
\label{Eq:2.8a}
\mathrm{ad}_{\Omega}(A) = \ri \varphi A.
\end{equation}
Moreover, by a well-known fact from Lie theory,
$\Omega$ is $K$-conjugate to some element $\Delta \in \mathfrak{t}$,
i.e. $\Omega = U^{\dagger}\Delta U$,
cf. \cite{broedie} and thus
\begin{equation}
\label{Eq:2.8b}
\mathrm{ad}_{\Delta}(UAU^{\dagger}) =
U\mathrm{ad}_{U^{\dagger}\Delta U}(A)U^{\dagger}. 
\end{equation}
Now, Eqs. (\ref{Eq:2.8a}) and (\ref{Eq:2.8b}) imply
$\mathrm{ad}_{\Delta}(UAU^{\dagger}) =  \ri \varphi UAU^{\dagger}$.
Hence, $UAU^{\dagger}$ is an eigenvector of $\mathrm{ad}_{\Delta}$
and thus the proof of (b) is complete.
\end{proof}

\begin{remark}
\label{Rem:2.3}
Equation (\ref{eq2.6}) in the  above proof---which can be rewritten
as
\begin{equation}
\e^{\ri k\varphi_0}A = \e^{\Omega_0 k}A\e^{-\Omega_0 k}
\end{equation}
for all $k \in \Z$ and  some $\Omega_0 \in \mathfrak{k}$---does in
general not imply
\begin{equation}\label{eq2.9}
\e^{\ri \varphi_0 t}A = \e^{\Omega_0 t}A\e^{-\Omega_0 t}
\end{equation}
for all $t \in \R$ as the following example shows. 
\end{remark}

\begin{example}
\label{ex5}
For $a,b \in \R$ let
\begin{equation*}
A :=
\begin{bmatrix}
a & b\\
b & -a
\end{bmatrix}
\quad\mbox{and}\quad
U_0 := 
\begin{bmatrix}
0 & 1 \\
-1 & 0
\end{bmatrix}
=
\e^{\Omega_0}
\quad\mbox{with}\quad
\Omega_0 :=
\frac{\pi}{2}
\begin{bmatrix}
0 & 1 \\
-1 & 0
\end{bmatrix}.
\end{equation*}
Then a straightforward computation yields
$U_0 A U_0^{\dagger} = - A$
and thus
$$
U_0^k A (U_0^k)^{\dagger} = 
\e^{\Omega k} A \e^{-\Omega k} = \e^{k \pi \ri} A
$$
for all $k \in \Z$. However, $A$ and $\Omega_0$ do not satisfy
Eq. (\ref{eq2.9}), as the necessary condition of Lemma \ref{lemrotsym}
is not met, i.e., $A$ is not an eigenvector of
$\mathrm{ad}_{\Omega_0}$ to the eigenvalue $-\ri\pi$.

\begin{problem}
Here, we have $\varphi_0 = \pi$, which is obvious a
rational multiple of $2\pi$ and thus does not completely
fit the previous requirements. Actually, we have the following
conjecture.
\end{problem}
 
\begin{conjecture}
It is impossible to find a counterexample also satisfing
$\varphi_0 \not\in 2\pi\Q$. To see this, one probably has to use
deeper results on the structure of the torus algebra and its
root space decomposition of $K$. 
\end{conjecture}
\end{example}

\noindent
Theorem \ref{Thm:rotsym} suggests that the set
\begin{equation*}
E(\mathfrak{t}) :=
\bigcup_{\Delta \in \mathfrak{t},\; \varphi \neq 0} E_{\varphi}(\Delta),
\end{equation*}
where $\mathfrak{t}$ is an arbitray torus algebra of $\mathfrak{k}$,
plays a crucial role in the further analysis of matrices with
rotationally symmetric relative $C$"~numerical range. More precisely,
we have the following corollary.

\begin{corollary}
\label{cor1rotsym}
Let $\mathfrak{t}$ be a torus algebra of $\mathfrak{k}$. The  relative
$C$"~numerical range of $A \in \C^{N \times N}$, $A \neq 0$ is 
rotationally symmetric for all $C \in \C^{N \times N}$ if and only if
there exists $U \in K$ such that $UAU^{\dagger}$ is in $E(\mathfrak{t})$.
\end{corollary} 

\begin{proof} 
This follows immediately from Theorem \ref{Thm:rotsym} and
Proposition \ref{Prop:rotsym}.
\end{proof}

\noindent
Next, we show that part (d) and (e) of Theorem \ref{thm:disc} can
easily be derived from Theorem \ref{Thm:rotsym} by choosing $K=U(N)$.
 
\begin{corollary}
\label{cor2rotsym}
The classical $C$"~numerical range of $A$ is rotationally symmetric
for all $C \in \C^{N \times N}$ if and only if $A$ is unitarily similar
to a block-shift matrix $M$, i.e. $M = (M_{kl})_{1 \leq k,l \leq m}$
is of block form such that all the $M_{kk}$ are square blocks and
$M_{kl} = 0$ if $l+1 \neq k$. 
\end{corollary} 

\begin{proof}
We may assume without loss of generality that $A \neq 0$. Applying
Theorem \ref{Thm:rotsym}(b) to $K := U(N)$ and the torus algebra 
$\mathfrak{t} :=
\{\Delta \in \mathfrak{u}(N) \;|\; \mbox{$\Delta$ diagonal}\}$
yields that the unitary orbit of $A$ is weakly rotationally symmetric
if and only if $A$ is unitarily similar to a matrix $M$ such that
\begin{equation}\label{eq2.10}
\mathrm{ad}_{\Delta}(M) = \ri \varphi M
\end{equation}
for some $\Delta \in \mathfrak{t}$ and $\varphi \in \R \setminus \{0\}$.
Now, $\Delta$ can be arranged such that
\begin{equation}\label{eq2.11}
\Delta =
\ri \cdot
\diag (
\underbrace{\lambda_1,\dots, \lambda_1}_{n_1-\mbox{times}},
\dots,
\underbrace{\lambda_m, \dots, \lambda_m}_{n_m-\mbox{times}}),
\quad
\sum_{j=1}^{m}n_j = n
\end{equation}
and 
\begin{equation}
\label{eq2.12}
\lambda_k - \lambda_{l} = \varphi
\quad\Longrightarrow\quad
k = l + 1
\end{equation}
for all $1 \leq k,l \leq m$. Choosing a block partition of $M$
corresponding to the one of $\Delta$, Eqs. (\ref{eq2.10}) and
(\ref{eq2.11}) yield
\begin{equation}
\label{eq2.13}
(\lambda_k - \lambda_l - \varphi)M_{kl} = 0
\end{equation}
for all $1 \leq k,l \leq m$. Thus condition (\ref{eq2.12}) implies
$M_{kl} = 0$ if  $k \neq l+1$ and hence $M$ has the required form.
\end{proof}

\begin{remark}
In contrast to part (f) of Theorem \ref{thm:disc} on the classical
$C$"~numerical range of $A$, the relative one need not be a circular disc
in order to be rotationally symmetric, a counterexample being provided
by Example \ref{ex1}.
\end{remark}

Finally, we want to obtain some information on the Lie-algebraic
structure of the set of all matrices with rotationally symmetric
relative $C$"~numerical range.

\begin{lemma}
\label{lemliestruc}
Let $\Omega$ be skew-Hermitian and let $A \neq 0$ be an eigenvector
of $\mathrm{ad}_{\Omega}$ to the non-trivial eigenvalue $\ri \varphi$,
$\varphi \in \R$. Then we have
\begin{enumerate}
\item 
$A^{\dagger}$ is an eigenvector of $\mathrm{ad}_{\Omega}$
to a non-trivial eigenvalue $-\ri \varphi$. 
\item
$A$ is nilpotent.
\item 
$[A,A^{\dagger}] \neq 0$ and $[\Omega,[A,A^{\dagger}]] = 0$.
\end{enumerate}
\end{lemma} 

\begin{proof}
(a) Let $\Omega$ be skew-Hermitian and $\varphi \neq 0$ such
that $\mathrm{ad}_{\Omega}(A) = \ri \varphi A$. Then
\begin{equation*}
\mathrm{ad}_{\Omega}(A^{\dagger}) =
\big(\mathrm{ad}_{\Omega}(A)\big)^{\dagger} =
-\ri \varphi A^{\dagger}.
\end{equation*}

\medskip

\noindent
(b) From the identity $[\Omega,A] = \ri \varphi A$ we obtain
$\Omega A^n - A^n \Omega= n\ri \varphi A^n$ for all $n \in \N$
by induction. Therefore, we have
\begin{equation}
\label{eq:nil}
n |\varphi| \|A\|^n \leq 2 \|\Omega\| \cdot \|A\|^n
\end{equation}
for all $n \in \N$, where $\|\cdot\|$ denotes the Frobenius norm.
This implies $A^n = 0$ for some $n \in \N$, otherwise Eq. (\ref{eq:nil})
would contradict the fact $\|\Omega\| < \infty$.

\medskip

\noindent
(c) Again, let $\Omega$ be skew-Hermitian and $\varphi \neq 0$
such that $\mathrm{ad}_{\Omega}(A) = \ri \varphi A$. Then
by the Jacobi-identity for the double commutator we obtain
\begin{eqnarray*}
[\Omega,[A,A^{\dagger}]]
& = &
- \Big([A,[A^{\dagger},\Omega]] + [A^{\dagger},[\Omega,A]] \Big)\\
& = &
[A,\mathrm{ad}_{\Omega}(A^{\dagger})] -
[A^{\dagger},\mathrm{ad}_{\Omega}(A)]\\
& = &
-\ri \varphi [A,A^{\dagger}] - \ri \varphi [A^{\dagger},A] = 0,
\end{eqnarray*}
i.e., $\Omega$ and $[A,A^{\dagger}]$ commute.  To prove that
$[A,A^{\dagger}]$ does not vanish, we assume the converse. Hence,
$A$ is normal and thus part (b) implies $A=0$.
This, however, contradicts our assumptions
on $A$ and therefore $[A,A^{\dagger}] \neq 0$.
\end{proof}

\begin{corollary}
\label{corliestruc}
\begin{enumerate}
\item 
The $K$-orbit of $A$ is weakly rotationally symmetric if and only if
the $K$-orbit of $A^{\dagger}$ is as well.
\item 
If the $K$-orbit of $A$ is weakly rotationally symmetric then also
the $K$-orbit of $[[A,A^{\dagger}],A]$.
\end{enumerate}
\end{corollary}

\begin{proof}
(a) This follows immediately by Lemma \ref{lemliestruc}(a) and
Theorem \ref{Thm:rotsym}(a).

\bigskip
\noindent
(b) Lemma \ref{lemliestruc} and the identity
$[\mathrm{ad}_{X},\mathrm{ad}_{Y}] = \mathrm{ad}_{[X,Y]}$ imply
that $\mathrm{ad}_{\Omega}$ and $\mathrm{ad}_{[A,A^{\dagger}]}$
commute and thus they admit a simultaneous eigenspace decomposition.
Hence $E_{\varphi}(\Omega)$ is invariant under
$\mathrm{ad}_{[A,A^{\dagger}]}$. In particular,
$\mathrm{ad}_{[A,A^{\dagger}]}(A) = [[A,A^{\dagger}],A]$ is
again an eigenvector of $\mathrm{ad}_{\Omega}$ to the eigenvalue 
$\ri \varphi$. Therefore, Theorem \ref{Thm:rotsym}(a) yields the
desired result. 
\end{proof}

\noindent 
So far we have seen that $A$ and $[A,A^{\dagger}]$ are contained in
$E_{\varphi}(\Omega)$. This, however, does not imply that
$[A,A^{\dagger}] = \lambda A$ for some $\lambda \in \C$. Therefore,
we introduce the following notion to analyse the situation in more
detail. For $A \in \C^{N \times N}$ let the \emph{separation index}
$I_s(A)$ of $A$ be defined by 
\begin{equation}
I_s(A) := \min\{\mathrm{dim}\; E_{\varphi}(\Omega) \;|\; 
A \in  E_{\varphi}(\Omega),\, \Omega \in \mathfrak{u}(N),\,
\varphi \in \R,\, \varphi \neq 0\}.
\end{equation}
If $A$ is not contained in any eigenspace $E_{\varphi}(\Omega)$,
then we set $I_s(A) := -\infty$.

\begin{proposition}
\label{Prop:liestruc}
If the separation index of $A$ is equal to $1$ then the Lie algebra
generated by $A-A^{\dagger}$, $\ri A + \ri A^{\dagger}$
and $\ri[A,A^{\dagger}]$ is isomorphic to $\mathfrak{su}(2)$.
\end{proposition} 

\begin{proof}
By assumption there exist $\Omega \in \mathfrak{u}(N)$ and
$\varphi \in \R$, $\varphi \neq 0$ such that $A \in E_{\varphi}(\Omega)$
with $\mathrm{dim}\; E_{\varphi}(\Omega) = 1$.
As in Corollary \ref{corliestruc}(b), we obtain the invariance of 
$E_{\varphi}(\Omega)$ under $\mathrm{ad}_{[A,A^{\dagger}]}$
and thus
\begin{equation*}
\mathrm{ad}_{[A,A^{\dagger}]}(A) = \lambda A
\end{equation*} 
for some $\lambda \in \R$. Here, $\lambda$ has to be real, since the
operator $\mathrm{ad}_{[A,A^{\dagger}]}$ is Hermitian with respect to
the scalar product $(A,C) \mapsto \tr (C^{\dagger}A)$. Moreover, as in
Lemma \ref{lemliestruc} we obtain
\begin{equation*}
\mathrm{ad}_{[A,A^{\dagger}]}(A^{\dagger}) = -\lambda A^{\dagger}.
\end{equation*} 
Hence, we have the following commutator relations:
\begin{eqnarray*}
[(A - A^{\dagger}),\ri(A + A^{\dagger})]
& = &
2\ri[A,A^{\dagger}],\\[2mm]
[\ri(A + A^{\dagger}),\ri[A,A^{\dagger}]]
& = &
\lambda (A - A^{\dagger}),\\[2mm]
[\ri[A,A^{\dagger}],(A - A^{\dagger})]
& = &
\lambda \ri(A + A^{\dagger}).
\end{eqnarray*}
Therefore, $X := \ri[A,A^{\dagger}]$, $Y := A - A^{\dagger}$ and
$Z := \ri A + \ri A^{\dagger}$ generate a $3$-dimensional Lie subalgebra
of $\mathfrak{u}(N)$. If $\lambda \neq 0$ then a straightforward
rescaling of $X,Y$ and $Z$ shows that the generated Lie subalgebra
is isomorphic to $\mathfrak{su}(2)$. If $\lambda = 0$ then $X$ and $Y$
as well as $X$ and $Z$ commute. Hence we can assume that $X$ and $Y$
are diagonal. This, however, contradicts $[Y,Z] = X$ and thus we
are done.
\end{proof}

\noindent
An alternative approach to Proposition \ref{Prop:liestruc} is given
by the following lemma, which explicitly determines all $A$ with
$I_s(A) = 1$. It is a straightforward consequence of Corollary
\ref{cor2rotsym} and therefore stated without proof. 

\begin{lemma}
\label{Lem:liestruc}
The separation index of $A$ is $1$ if and only if $A$ is unitarily similar
to $\lambda E_{ij}$ for some $\lambda \in \C$, $\lambda \neq 0$, where all
entries of $E_{ij}$ are zero except the one in the $i$-th row and
$j$-th column with $1 \leq i,j \leq N$, $i \neq j$.
\end{lemma}

\begin{remark}
\label{Rem:4}
\begin{enumerate}
\item 
Proposition \ref{Prop:liestruc} is independent of the subgroup
$K \subset U(N)$. In particular, $A$ can be the eigenvector of
some $\Omega \in \mathfrak{k}$ with
$I_s(A) = 1$, while $\ri[A,A^{\dagger}]$ is not contained in
the Lie algebra $\mathfrak{k}$ of $K$.
For instance, let 
$$
K:=
\left\{
\begin{bmatrix}
U & 0\\
0 & U
\end{bmatrix}
\;\Big|\;
U \in SU(2)
\right\},
$$ 
$$
\Omega :=
\begin{bmatrix}
\ri & 0 & 0 & 0\\
0 & -\ri & 0 & 0\\
0 & 0 & \ri & 0\\
0 & 0 & 0 & -\ri
\end{bmatrix}
\quad\mbox{and}\quad
A :=
\begin{bmatrix}
0 & 0 & 0 & 0\\
1 & 0 & 0 & 0\\
0 & 0 & 0 & 0\\
0 & 0 & 0 & 0
\end{bmatrix},
\quad\mbox{but}\quad
[A,A^{\dagger}] =
\begin{bmatrix}
-1 & 0 & 0 & 0\\
0 & 1 & 0 & 0\\
0 & 0 & 0 & 0\\
0 & 0 & 0 & 0
\end{bmatrix}.
$$
Thus the subgroup corresponding to the subalgebra generated
by $A-A^{\dagger}$, $\ri A + \ri A^{\dagger}$ and $\ri[A,A^{\dagger}]$ 
is in general not contained in $K$, even if the $K$-orbit of
$A$ is weakly rotationally symmetric. 
On the other hand, characterizing all subgroups $K$ having
the inclusion property
\begin{equation*}
A-A^{\dagger},\, \ri A + \ri A^{\dagger},\, \ri[A,A^{\dagger}]\,
\in \mathfrak{k} 
\end{equation*}
for all $A$ with weakly rotationally symmetric $K$-orbit seems to be
an open problem.
\item 
If the separation index of $A$ is greater than $1$, then Proposition
\ref{Prop:liestruc} is in general not true, as the following example
shows.
$$
A :=
\begin{bmatrix}
0 & 0 & 0 & 0\\
1 & 0 & 0 & 0\\
0 & 0 & 0 & 0\\
0 & 0 & 2 & 0
\end{bmatrix}
\quad\mbox{and}\quad
[A,A^{\dagger}] :=
\begin{bmatrix}
-1 & 0 & 0 & 0\\
0 & 1 & 0 & 0\\
0 & 0 & -4 & 0\\
0 & 0 & 0 & 4
\end{bmatrix}.
$$
By Corollary \ref{cor2rotsym}, we have $I_s(A)=2$. Moreover,
the Lie algebra generated by $A-A^{\dagger}$,
$\ri A + \ri A^{\dagger}$ and $\ri[A,A^{\dagger}]$ is not
isomorphic to $\mathfrak{su}(2)$, as claimed above.
\end{enumerate}
\end{remark}


\section{The Local $C$"~Numerical Range}
\label{sec:3}

In this subsection we specify the previous results to the $n$-fold
tensor product of $SU(2)$, i.e.
\begin{equation*}
K := SU_{\rm loc}(2^n) :=
\underbrace{SU(2) \otimes \dots \otimes SU(2)}_{\mbox{$n$-times}}
\subset SU(2^N).
\end{equation*}

\noindent
In quantum mechanics and, in particular, in quantum information,
$SU_{\rm loc}(2^n)$ is called the \emph{subgroup} of \emph{local action}.
Therefore, we call the corresponding relative $C$"~numerical range
the \emph{local} $C$\emph{-numerical range of} $A$ and introduce the
short-hand notation
\begin{equation}
\label{Eq:2.20}
W_{\mathrm{loc}}(C,A) := W_{SU_{\rm loc}(2^n)}(A,C).
\end{equation}
Note that replacing $SU_{\rm loc}(2^n)$ by
$U(2) \otimes \dots \otimes U(2)$ in Definition (\ref{Eq:2.20})
would yield the same local $C$"~numerical range, 
which can easily be seen by the identity
$$
(\e^{\ri \varphi_1}U_1) \otimes \dots \otimes (\e^{\ri \varphi_n}U_n)
= \e^{\ri \varphi_1 + \dots \varphi_n} (U_1 \otimes \dots \otimes U_n).
$$

\begin{remark}
\label{rem:3.1}
Following Eq.~(\ref{num-Cnum}) one might naively assign a relative
numerical range $W_K(A)$ to an
operator $A \in \C^{N \times N}$ by a definition like
$W_K(A):=W_K(xx^{\dagger},A)$ with $x \in \C^N$, $\|x\|_2 = 1$.
However, such a concept is inappropriate as it would
depend on the particular choice of $x \in \C^N$. Yet, for the local
case or more general, if $K = SU(N_1) \otimes \dots \otimes SU(N_n)$ 
is a tensor product of special unitary groups,
there is a canonical subset of the unit sphere, to wit the set of all
$x=x_1 \otimes \dots \otimes x_n$ with $x_k \in \C^{N_k}$, $\|x_k\|=1$,
on which $K$ acts transitively.  This allows for properly defining the
\emph{local numerical range of} $A$ as the set
\begin{equation*}
W_{\rm loc}(A)
:=\big\{\tr(x^{\dagger}Ax) \;\big|\; x =x_1 \otimes \dots \otimes x_n,
x_k \in \C^{2}, \|x_k\|=1 \big\},
\end{equation*}
which in turn immediately yields the local analogue of Eq.~(\ref{num-Cnum}).
Moreover, 
note that in physical terms, the local
numerical range is nothing else than the classical numerical range
restriced to the set of all \emph{pure product states}. Some of its
implications are analysed in the accompanying paper \cite{schuldihegla:07}.
\end{remark}

Now, for applying Theorem \ref{Thm:rotsym}, we have to choose a torus
algebra in the Lie algebra of $SU_{\rm loc}(2^N)$. A straighforward way
of doing so is presented in the following.
Let $K_1 \subset \C^{N_1 \times N_1}$ and $K_2 \subset \C^{N_2 \times N_2}$
be Lie subgroups with Lie algebras $\mathfrak{k}_1$ and $\mathfrak{k}_2$,  
respectively. Then the Lie algebra of the tensor product $K_1 \otimes K_2$ is
denoted by $\mathfrak{k}_1 \widehat{\oplus} \mathfrak{k}_2$. It
is given by
\begin{equation}
\mathfrak{k}_1 \widehat{\oplus} \mathfrak{k}_2 :=
\{\Omega_1 \otimes I_{N_2} + I_{N_1} \otimes \Omega_2 \;|\;
\Omega_1 \in \mathfrak{k}_1,\, \Omega_2 \in \mathfrak{k}_2\}
\subset \C^{N_1N_2 \times N_1N_2 }.
\end{equation}
Moreover, let $sl_{\C}(N)$ denote the set of
all $A \in \C^{N \times N}$ with $\tr A = 0$. 

\begin{lemma}
\label{lemlocrange}
Let $\mathfrak{t}_1$ and  $\mathfrak{t}_2$ be  torus algebras of
the subalgebras $\mathfrak{k}_1 \subset \C^{N_1 \times N_1}$ and
$\mathfrak{k}_2 \subset \C^{N_2 \times N_2}$, respectively.
Then $\mathfrak{t}_1 \widehat{\oplus} \mathfrak{t}_2$ is a torus
algebra of $\mathfrak{k}_1 \widehat{\oplus} \mathfrak{k}_2$.
If, moreover, $\mathfrak{k}_1 \subset sl_{\C}(N_1)$
and $\mathfrak{k}_2 \subset sl_{\C}(N_2)$,
then the converse is also true.
\end{lemma}

\begin{proof} 
Let $\Omega = \Omega_1 \otimes I_{N_2} + I_{N_1} \otimes \Omega_2
\in \mathfrak{k}_1 \widehat{\oplus} \mathfrak{k}_2$ such that
$[\Omega, \Omega'] = 0$ for all
$\Omega' \in \mathfrak{t}_1 \widehat{\oplus} \mathfrak{t}_2$,
i.e.
\begin{eqnarray*}
\lefteqn{[\Omega_1 \otimes I_{N_2} + I_{N_1} \otimes \Omega_2,
\Omega'_1 \otimes I_{N_2} + I_{N_1} \otimes \Omega'_2] =}\\
& = & 
[\Omega_1,\Omega'_1] \otimes I_{N_2} + 
I_{N_1} \otimes [\Omega_2,\Omega'_2] = 0
\end{eqnarray*}
for all $\Omega'_1 \in \mathfrak{t}_1$ and
$\Omega'_2 \in \mathfrak{t}_2$. By the fact that
$[\Omega_1,\Omega'_1] \otimes I_{N_2}$ and 
$I_{N_1} \otimes [\Omega_2,\Omega'_2]$ are orthogonal
with respect to the scalar product $(A,C) \mapsto \tr (C^{\dagger}A)$,
we obtain the equivalence
\begin{eqnarray*}
\label{eq2.25}
& 
[\Omega, \Omega'] = 0 \quad\mbox{for all
$\Omega' \in \mathfrak{t}_1 \widehat{\oplus} \mathfrak{t}_2$}
&\\
&
\Longleftrightarrow
&\\
&[\Omega_1,\Omega'_1] = 0
\quad\mbox{and}\quad
[\Omega_2,\Omega'_2] = 0
\quad\mbox{for all $\Omega'_1 \in \mathfrak{t}_1$,
$\Omega'_2 \in \mathfrak{t}_2$}
&
\end{eqnarray*}
Now, if $\mathfrak{t}_1$ and $\mathfrak{t}_2$ are maximal Abelian,
then $\Omega_1$ and $\Omega_2$ are contained in $\mathfrak{t}_1$ and
$\mathfrak{t}_2$, respectively, and thus
$\mathfrak{t}_1 \widehat{\oplus} \mathfrak{t}_2$ is maximal Abelian,
too. On the other hand, let $\mathfrak{t}_1, \mathfrak{t}_2$ be Abelian and
suppose maximality of $\mathfrak{t}_1 \widehat{\oplus} \mathfrak{t}_2$,
the above equivalence shows that
$\Omega_1 \otimes I_{N_2} + I_{N_1} \otimes \Omega_2$ belongs
to $\mathfrak{t}_1 \widehat{\oplus} \mathfrak{t}_2$, if
$[\Omega_i,\Omega'_i] = 0$ for all $\Omega'_i \in \mathfrak{t}_i$ and
$i = 1,2$. Hence, it follows $\Omega_1 \in \mathfrak{t}_1$ and
$\Omega_2 \in \mathfrak{t}_2$, if the map
\begin{equation}\label{eq2.27}
(\Omega_1,\Omega_2) \mapsto
\Omega_1 \otimes I_{N_2} + I_{N_1} \otimes \Omega_2
\end{equation}
is one-to-one. This, however, is guaranteed under the additional
assumption $\mathfrak{k}_1 \subset sl_{\C}(N_1)$ and
$\mathfrak{k}_2 \subset sl_{\C}(N_2)$. Therefore, $\mathfrak{t}_1$
and $\mathfrak{t}_2$ are maximal Abelian, too.
\end{proof}

\noindent
Note that the additional assumption for the converse in Lemma
\ref{lemlocrange} is necessary as the following example shows.

\begin{example}
Let $\mathfrak{k}_1 := \mathfrak{k}_2 := \mathfrak{u}(2)$
and define 
$$
\mathfrak{t}_1 :=
\left\{
\begin{bmatrix}
\ri \lambda & 0\\
0 & -\ri \lambda
\end{bmatrix}
\;\Big|\; \lambda \in \R
\right\}
\quad\mbox{and}\quad
\mathfrak{t}_2 :=
\left\{
\begin{bmatrix}
\ri \lambda & 0\\
0 & \ri \mu
\end{bmatrix}
\;\Big|\; \lambda,\mu \in \R
\right\}.
$$
Then $\mathfrak{t}_1 \widehat{\oplus} \mathfrak{t}_2$ is a torus
algebra in $\mathfrak{u}(2)\widehat{\oplus}\mathfrak{u}(2)$.
However, $\mathfrak{t}_1$ is not maximal Abelian in
$\mathfrak{k}_1 = \mathfrak{u}(2)$.
\end{example}

Now, let $\mathfrak{su}_{\rm loc}(2^n)$ be the Lie algebra of
$SU_{\rm loc}(2^n)$ and let
$\mathfrak{t}_{\mathrm{loc}} \subset \mathfrak{su}_{\rm loc}(2^n)$
be the subset of all diagonal matrices. Obviously,
$\mathfrak{t}_{\mathrm{loc}}$ is a torus algebra of
$\mathfrak{su}_{\rm loc}(2^n)$ by Lemma
\ref{lemlocrange} which yields the following corollary.

\begin{corollary}
\label{corlocrange}
The local $C$"~numerical range $W_{\mathrm{loc}}(C,A)$ of
$A\in \C^{2^n \times 2^n}$ is rotationally symmetric for all
$C \in \C^{2^n \times 2^n}$ if and only
if there exists $U \in SU_{\rm loc}(2^n)$ such that
$UAU^{\dagger} \in E(\mathfrak{t}_{\mathrm{loc}})$, i.e. 
\begin{equation}
\label{eq2.30}
[\Delta, UAU^{\dagger}] =  \ri \varphi UAU^{\dagger}.
\end{equation}
for some $\Delta \in \mathfrak{t}_{\mathrm{loc}}$ and $\varphi \in \R$,
$\varphi \neq 0$.
\end{corollary}

\begin{proof}
This follows immediately from Corollary \ref{cor1rotsym} and
Lemma \ref{lemlocrange}.
\end{proof}

Finally, we are prepared to present the main result of this section,
which roughly speaking excludes the possibility of an annulus for
rotationlly symmetric local $C$"~numerical ranges.

\begin{theorem}
\label{thmlocrange}
The local $C$"~numerical range $W_{\mathrm{loc}}(C,A)$ of
$A \in \C^{2^n \times 2^n}$ is rotationally symmetric for all
$C \in \C^{2^n \times 2^n}$ if and only if it is a circular disc
in the complex plane centered at the origin
for all $C \in \C^{2^n \times 2^n}$.
\end{theorem}

\noindent
Before approaching Theorem \ref{thmlocrange} we provide the following
technical lemma.

\begin{lemma}
\label{lem3locrange}
Let $\Delta \in \mathfrak{t}_\mathrm{loc}$ and
$A = (a_{ij}) \in \C^{2^n \times 2^n}$
satisfy the relation
\begin{equation}\label{eq2.45}
[\Delta,A] = \ri \varphi A
\quad \mbox{for some $\varphi \in \Q$}.
\end{equation}
Then there exists a rational
$\Delta' \in \mathfrak{t}_\mathrm{loc}$ 
such that Eq. (\ref{eq2.45}) holds.
\end{lemma}

\begin{proof}
Let $\Delta \in \mathfrak{t}_\mathrm{loc}$, i.e.
\begin{eqnarray*}
\Delta
& = & 
\sum_{j=1}^{n}
I_2 \otimes \dots \otimes I_2 \otimes
\underbrace{
\begin{bmatrix}
\ri \lambda_j & 0\\
0 & -\ri \lambda_j
\end{bmatrix}}_{\mbox{$j$-th position}}
\otimes I_2 \otimes \dots \otimes I_2
\end{eqnarray*}
and let $\mu := (\mu_1, \dots, \mu_{2^n})^{\top}$ denote the diagonal
entries of $\Delta$, i.e. 
$\Delta = \ri \cdot \diag(\mu_1, \dots, \mu_{2^n}).$
Then one can find a matrix
$X_{\mathrm{loc}} \in \Q^{(2^n - n) \times 2^n}$ such that
\begin{equation}\label{eq2.46}
\Delta \in \mathfrak{t}_\mathrm{loc}
\quad\Longleftrightarrow\quad
X_{\mathrm{loc}} \, \mu = 0,
\end{equation}
cf. Lemma \ref{lemlocrange}. Moreover, a straightforward calculation
shows that
\begin{equation}\label{eq2.47}
[\Delta,A] = \ri \varphi A 
\quad\Longleftrightarrow\quad
\left\{
\begin{array}{l}
a_{ii} = 0 \quad \mbox{for all $i = 1, \dots, 2^n$}\\
X_{\mathrm{ad}} \, \mu = (\varphi,\dots,\varphi)^{\top},
\end{array}
\right.
\end{equation}
where $X_{\mathrm{ad}}$ is a matrix of appropriate size depending on $A$
with entries equal to $\pm 1$ or $0$. In particular, 
$X_{\mathrm{ad}} \in  \Q^{m \times 2^n}$.
Hence, we have
\begin{equation}\label{eq2.49}
[\Delta,A] = \ri \varphi A, \quad \Delta \in \mathfrak{t}_\mathrm{loc}
\quad\Longleftrightarrow\quad
\left\{
\begin{array}{l}
a_{ii} = 0 \quad \mbox{for all $i = 1, \dots, 2^n$}\\
X_{\mathrm{loc}}\, \mu = 0 \\
X_{\mathrm{ad}}\, \mu = (\varphi,\dots,\varphi)^{\top}
\end{array}
\right.
\end{equation}
with 
$\begin{bmatrix}
X_{\mathrm{loc}}^{\top} & X_{\mathrm{ad}}^{\top}
\end{bmatrix}
\in  \Q^{2^n \times ((2^n -n) + m)}$. 
Now, by assumption there exists a $\mu$ such that Eq.
(\ref{eq2.49}) is satisfied for some $\varphi \in \Q$.
This, however, implies that Eq. (\ref{eq2.49}) has in
particular rational solutions, i.e. solutions in $ \Q^{2^n}$. 
\end{proof}

\begin{proof}[of Theorem \ref{thmlocrange}]
``$\Longleftarrow$'': $\checkmark$

\medskip

\noindent
``$\Longrightarrow$'': Suppose that $W_{\mathrm{loc}}(C,A)$ is rotationally
symmetric for all $C \in \C^{2^n \times 2^n}$. It is sufficient to
show that zero is contained in $W_{\mathrm{loc}}(C,A)$. Therefore,
we can assume $\tr (C^{\dagger}A) \neq 0$ without loss of generality.
Thus, by Proposition \ref{Prop:rotsym} and Theorem \ref{Thm:rotsym}(a)
there exists $\Omega \in \mathfrak{su}_{\rm loc}(2^n)$ such that
\begin{equation*}
t \mapsto \omega(t) := \tr (C^{\dagger} \e^{\Omega t} A \e^{-\Omega t})
= \e^{\ri \varphi t} \tr (C^{\dagger}A),
\quad t \in \R
\end{equation*} 
is a circle around the origin in the complex plane. By Theorem
\ref{Thm:rotsym}(b) and Lemma \ref{lemlocrange} we can assume that
$\Omega$ is of diagonal form
\begin{equation}\label{eq2.30a2.50}
\Omega =
\sum_{j=1}^{N}
I_2 \otimes \dots \otimes I_2 \otimes
\underbrace{
\begin{bmatrix}
\ri \lambda_j & 0\\
0 & -\ri \lambda_j
\end{bmatrix}}_{\mbox{$j$-th position}}
\otimes I_2 \otimes \dots \otimes I_2.
\end{equation}
and satisfies the relation $[\Omega,A] = \ri \varphi A$. By rescaling
$\Omega$ such that $\varphi$ is rational and by invoking Lemma
\ref{lem3locrange} we further suppose that for $j = 1, \dots, n$
all $\lambda_j$ are rational.
Now, let $m$ be the least common multiple of the denominaters
of all $\lambda_j$ for $j = 1, \dots, N$. Then $m \varphi \in \Z$ and thus
$\omega|_{[0,2m\pi]}$ is a circle in the complex plane surrounding the origin
$(m \varphi)$-times. Moreover, we have
\begin{equation}\label{eq2.32}
\e^{2m\pi \Omega} = I_{2^n}.
\end{equation}
Therefore, the homotopy
$H:[0,2m\pi] \times [0,\T\frac{\pi}{2}] \to SU_{\rm loc}(2^n)$
of the form $H(t,s) = U(s)^{\dagger} \e^{t \Omega} U(s)$ with
$$
U(s) :=
\begin{bmatrix}
\cos s & \sin s\\
-\sin s & \cos s 
\end{bmatrix}
\otimes \dots \otimes
\begin{bmatrix}
\cos s & \sin s\\
-\sin s & \cos s 
\end{bmatrix}
$$
satisfies
\begin{equation}\label{eq2.34}
H(t,0) = \e^{t \Omega},
\quad
H(t,\T\frac{\pi}{2}) = \e^{-t \Omega}
\quad\mbox{and}\quad
H(0,s) = H(2m\pi,s) = I_{2^n}
\end{equation}
for all $(t,s) \in [0,2m\pi] \times [0,\T\frac{\pi}{2}]$. It follows
that $\omega|_{[0,2m\pi]}$ is homotopic to its inverse by the homotopy 
$$
h(t,s) := \tr \big(C^{\dagger} H(s,t)^{\dagger} A H(t,s)\big).
$$
Hence, $h$ has to cross the origin, cf. Appendix \ref{appendix},
Lemma \ref{lem:homotopy}, and thus, the origin is contained in
$W_{\mathrm{loc}}(C,A)$. Therefore, $W_{\mathrm{loc}}(C,A)$
is a circular disc.
\end{proof}

In the remainder of this section, we exemplify the previous results
by determining the set of all matrices $A \in \C^{4 \times 4}$, the local
$C$"~numerical range of which is a cirular disc centered at the origin.
These investigations will lead to a conjecture about ``local'' similarity to
block-shift form. But first---as promised---we tackle
the problem of computing all $A \in \C^{4 \times 4}$ with circular local
$C$"~numerical range. By Corollary \ref{corlocrange}, we can focus on
the set of all  $\widehat{A} \in \C^{4 \times 4}$ which satisfy
\begin{equation}
\label{eq2.35}
[\Delta, \widehat{A}] =  \ri \varphi \widehat{A}.
\end{equation}
for some $\Delta \in \mathfrak{t}_{\mathrm{loc}}$ and $\varphi \neq 0$.
Let $\widehat{A}:= (\hat{a}_{kl})$ and $\Delta :=
\diag(\lambda_1, \dots, \lambda_4) \in \mathfrak{t}_{\mathrm{loc}}$.
Then Eq. (\ref{eq2.35}) can be rewritten as 
\begin{equation}\label{eq2.36}
(\lambda_k - \lambda_l)\hat{a}_{kl} =  \varphi \hat{a}_{kl}
\end{equation}
for all $k,l = 1, \dots, 4$. Moreover, a straightforward calculation
shows 
\begin{equation}\label{eq2.37}
\Delta \in \mathfrak{t}_{\mathrm{loc}}
\quad\Longleftrightarrow\quad
\Delta =
\begin{bmatrix}
\ri \lambda& 0 & 0 & 0\\
0 & \ri \mu & 0 & 0\\
0 & 0 & -\ri \mu & 0\\
0 & 0 & 0 & -\ri \lambda
\end{bmatrix}
\quad \mbox{with $\lambda, \mu \in \R$}.
\end{equation}
Now, considering all possibilities for $\lambda_k - \lambda_l$ to
be equal to $\varphi \neq 0$ and taking into account Eq. (\ref{eq2.37})
leads to 32 different cases.
However, by the symmetry of Eq. (\ref{eq2.36}), these can be reduced to
the following 16 ones, while the remaining ones 
can be obtained by transposition. 

\begin{center}
\begin{tabular}{c|c|c|c}
Case & Eigenvalue & Case & Eigenvalue\\[1mm]
\hline
1 & $\varphi = \mu - \lambda$ &
9 & $\varphi = \mu -\lambda = -2\mu$ \\[1mm]
2 & $\varphi = -\mu - \lambda$ &
10 & $\varphi = \mu - \lambda = 2\mu$ \\[1mm]
3 & $\varphi = -2\lambda$ &
11 & $\varphi = -\mu - \lambda = -2\lambda$ \\[1mm]
4 & $\varphi = -2\mu$ &
12 & $\varphi = -\mu - \lambda = 2\lambda$ \\[1mm]
5 & $\varphi = \mu - \lambda = -\mu - \lambda$ &
13 &  $\varphi = -\mu - \lambda = -2\mu$ \\[1mm]
6 & $\varphi = \mu - \lambda = \mu + \lambda$ &
14 & $\varphi = -\mu - \lambda = 2\mu$ \\[1mm]
7 & $\varphi = \mu - \lambda = -2\lambda$ &
15 & $\varphi = -2\lambda = -2\mu$\\[1mm]
8 & $\varphi = \mu - \lambda = +2\lambda$ &
16 & $\varphi = -2\lambda = 2\mu$\\[1mm]
\end{tabular}
\end{center}

\noindent
For example, Case 1 in the above table means that
$\lambda_2 - \lambda_1 = \lambda_4 - \lambda_3
= \mu - \lambda = \varphi \neq 0$
and all other differences $\lambda_k - \lambda_l$ do not equal
$\varphi$. Hence, Eq. (\ref{eq2.36}) reads
$\varphi\hat{a}_{21} =  \varphi \hat{a}_{21}$,
$\varphi\hat{a}_{43} =  \varphi \hat{a}_{43}$,
and $\hat{a}_{kl} = 0$ otherwise.
Thus $\widehat{A}$ has the form
$$
\mbox{Case 1:}\;
\begin{bmatrix}
0 & 0 & 0 & 0\\
* & 0 & 0 & 0\\
0 & 0 & 0 & 0\\
0 & 0 & * & 0
\end{bmatrix},
$$
where the symbol $*$ denotes an arbitrary complex number.
In the same way one can compute $\widehat{A}$ in all other cases.
Here, we only list the solutions for the above table. The remaining
ones---as mentioned before---can be obtained by transposition. 
\begin{eqnarray*}
\mbox{Case 2:}\;
\begin{bmatrix}
0 & 0 & 0 & 0\\
0 & 0 & 0 & 0\\
* & 0 & 0 & 0\\
0 & * & 0 & 0
\end{bmatrix},
&
\mbox{Case 3:}\;
\begin{bmatrix}
0 & 0 & 0 & 0\\
0 & 0 & 0 & 0\\
0 & 0 & 0 & 0\\
* & 0 & 0 & 0
\end{bmatrix},
&
\mbox{Case 4:}\;
\begin{bmatrix}
0 & 0 & 0 & 0\\
0 & 0 & 0 & 0\\
0 & * & 0 & 0\\
0 & 0 & 0 & 0
\end{bmatrix},\\\\
\mbox{Case 5:}\;
\begin{bmatrix}
0 & 0 & 0 & 0\\
* & 0 & 0 & 0\\
* & 0 & 0 & 0\\
0 & * & * & 0
\end{bmatrix},
&
\mbox{Case 6:}\;
\begin{bmatrix}
0 & 0 & * & 0\\
* & 0 & 0 & *\\
0 & 0 & 0 & 0\\
0 & 0 & * & 0
\end{bmatrix},
&
\mbox{Case 7:}\;
\begin{bmatrix}
0 & 0 & 0 & 0\\
* & 0 & 0 & 0\\
0 & 0 & 0 & 0\\
* & 0 & * & 0
\end{bmatrix},\\\\
\mbox{Case 8:}\;
\begin{bmatrix}
0 & 0 & 0 & *\\
* & 0 & 0 & 0\\
0 & 0 & 0 & 0\\
0 & 0 & * & 0
\end{bmatrix},
&
\mbox{Case 9:}\;
\begin{bmatrix}
0 & 0 & 0 & 0\\
* & 0 & 0 & 0\\
0 & * & 0 & 0\\
0 & 0 & * & 0
\end{bmatrix},
&
\mbox{Case 10:}\;
\begin{bmatrix}
0 & 0 & 0 & 0\\
* & 0 & * & 0\\
0 & 0 & 0 & 0\\
0 & 0 & * & 0
\end{bmatrix}\\\\
\mbox{Case 11:}\;
\begin{bmatrix}
0 & 0 & 0 & 0\\
0 & 0 & 0 & 0\\
* & 0 & 0 & 0\\
* & * & 0 & 0
\end{bmatrix},
&
\mbox{Case 12:}\;
\begin{bmatrix}
0 & 0 & 0 & *\\
0 & 0 & 0 & 0\\
* & 0 & 0 & 0\\
0 & * & 0 & 0
\end{bmatrix},
&
\mbox{Case 13:}\;
\begin{bmatrix}
0 & 0 & 0 & 0\\
0 & 0 & 0 & 0\\
* & * & 0 & 0\\
0 & * & 0 & 0
\end{bmatrix}\\\\
\mbox{Case 14:}\;
\begin{bmatrix}
0 & 0 & 0 & 0\\
0 & 0 & * & 0\\
* & 0 & 0 & 0\\
0 & * & 0 & 0
\end{bmatrix},
&
\mbox{Case 15:}\;
\begin{bmatrix}
0 & 0 & 0 & 0\\
0 & 0 & 0 & 0\\
* & * & 0 & 0\\
* & * & 0 & 0
\end{bmatrix},
&
\mbox{Case 16:}\;
\begin{bmatrix}
0 & 0 & 0 & 0\\
* & 0 & * & 0\\
0 & 0 & 0 & 0\\
* & 0 & * & 0
\end{bmatrix}.
\end{eqnarray*}

Finally, we are prepared to provide a rigorous proof of what we claimed
in Example \ref{ex4}: The local $C$"~numerical range of the block-shift
matrix
$$
A := 
\begin{bmatrix}
0 & 0 & 0 & 0\\
1 & 0 & 0 & 0\\
1 & 0 & 0 & 0\\
1 & 0 & 0 & 0
\end{bmatrix}
$$
is not rotationally symmetric for all $C \in \C^{4 \times 4}$.
According to the above classification, we have to show that $A$ is not
locally unitarily similar, i.e. similar via an element in
$SU(2) \otimes SU(2)$, to one of the above cases.
This, however, can be checked by ``brute force'' and is left to the reader.

\bigskip

The above computations reveal an interesting contrast to Corollary
\ref{cor2rotsym}: on the one hand, we have seen by Example \ref{ex4}
that not every matrix which is similar to block-shift from via a local
unitary transformation has a circular local $C$"~numerical range. On the
other hand, there are matrices with circular local $C$"~numerical range,
which are not locally unitarily similar to block-shift form, cf. Case 16
in the above table.
However, combining Corollary \ref{Cor:rotsym} and \ref{cor2rotsym},
every matrix with circular local $C$"~numerical range
has to be ``globally'' unitarily similar to block-shift form, i.e.
via a transformation in $U(2^n)$.
This raises the question: what is a smallest
subgroup $K'$ of $U(2^n)$ containing $SU_{\rm loc}(2^n)$ such that every
matrix with circular local $C$"~numerical range is similar to block-shift
form via a unitary transformation in $K'$? 
By Corollary \ref{corlocrange}, we can reduce the problem to studying
the smallest subgroup $\Pi'$ of (signed)
permutations, such that any element in $E(\mathfrak{t}_{\mathrm{loc}})$
is similar to block-shift form via a permutation in $\Pi'$.

One should note that in the above table all matrices are either in
block-shift form or similar to block-shift form via a permutation
of the following type:
\begin{equation*}
P_{\rm loc} := P_1 \otimes P_2,
\quad
P_1, P_2 \in
\left\{\pm I_2,
\pm
\begin{bmatrix}
0 & 1\\ 
-1 & 0
\end{bmatrix}
\right\}
\quad\mbox{or}\quad 
P_{\rm out} := 
\begin{bmatrix}
1 & 0 & 0 & 0\\
0 & 0 & 1 & 0\\
0 & -1 & 0 & 0\\
0 & 0 & 0 & 1
\end{bmatrix}.
\end{equation*}
This leads to the following conjecture:

\begin{conjecture}\label{conlocrange}
Every element of $E(\mathfrak{t}_\mathrm{loc})$ is similar to a
block-shift matrix via an element of
$\Pi_{\mathrm{loc}}^{\mathrm{ex}} :=
\Pi_{\mathrm{loc}} \cdot \Pi_{\mathrm{out}}$. 
Here, $\Pi_{\rm loc}$ denotes the subgroup of
all (signed) \emph{local} permutations,
i.e. $\Pi_{\rm loc}$ consists of all matrices of the form
$P_1 \otimes \dots \otimes P_n$ with
\begin{equation}\label{eq2.40}
P_k \in
\left\{\pm I_2,
\pm
\begin{bmatrix}
0 & 1\\ 
-1 & 0
\end{bmatrix}
\right\}
\end{equation}
for $k = 1, \dots, n$, and $\Pi_{\mathrm{out}}$ stands for
the subgroup generated by
\begin{equation*}
I_2 \otimes \dots \otimes I_2
\otimes P_{\rm out} \otimes
I_2 \otimes \dots \otimes I_2 \in U(2^n),
\quad  n \geq 2,
\end{equation*}
where the factor $P_{\rm out}$ appears in all possible positions. 
\end{conjecture}

\begin{remark}
\begin{enumerate}
\item 
Note that the two subgroups $\Pi_{\rm loc}$ and $\Pi_{\rm out}$ operate
via similarity on the set of tensor products
$A_1 \otimes \cdots \otimes A_n$ with $A_1, \dots, A_n \in \C^{2 \times 2}$
in a completely different way. While $\Pi_{\rm loc}$ acts on each factor
$A_k$ separately, $\Pi_{\rm out}$ does not effect the factors themselves but
interchanges their order. Moreover, $\Pi_{\rm loc}$ and $\Pi_{\rm out}$
commute and hence
$\Pi^{ex}_{\rm loc} := \Pi_{\rm loc} \cdot \Pi_{\rm out} =
\Pi_{\rm out} \cdot \Pi_{\rm loc}$
is isomorphic to the direct product of $\Pi_{\rm loc}$ and $\Pi_{\rm out}$.
We call it the \emph{extended local permutation group}.
It is easy to check that 
$E(\mathfrak{t}_\mathrm{loc})$ is invariant under conjugation
with elements of $\Pi_{\mathrm{loc}}^{\mathrm{ex}}$. Therefore,
we expect $\Pi_{\mathrm{loc}}^{\mathrm{ex}}$ to be the smallest
subgroup of permutations satisfying the above conjecture.
\item 
The group $\Pi_{\mathrm{loc}}^{\mathrm{ex}}$ is closely related
to the \emph{Weyl group} of $SU_{\rm loc}(2^n)$. More precisely,
the action of $\Pi_{\mathrm{loc}}$ on $\mathfrak{t}_{\rm loc}$
coincides with the Weyl group action of $SU_{\rm loc}(2^n)$
on the torus algebra $\mathfrak{t}_{\rm loc}$.
However, conjugation by elements of $\Pi_{\mathrm{out}}$ cannot be
achieved by elements of the Weyl group. These ideas suggest how to
generalise the above result to the settings of arbitrary compact
Lie groups.
\end{enumerate}
\end{remark}


\section{Conclusions}
\label{sec:4}

We introduced a new mathematical object, the \emph{relative}
$C$"~\emph{numerical range} $W_K(C,A)$ of an operator $A$.  
In particular, we studied its geometry by comparing its properties
with the classical $C$"~\emph{numerical range}. We showed that although
the {\em relative} $C$"~\emph{numerical range} is compact
and connected as in the classical case, it is neither star-shaped nor
simply connected. Moreover, necessary and
sufficient conditions for circular symmetry of $W_K(C,A)$ have been derived.
These results generalise a former theorem by Li and Tsing \cite{li:91} and
lead also to a deeper understanding of the classical case in Lie theoretical
terms. 
Moreover, in view of applications in quantum information, we introduced 
the \emph{local} $C$"~\emph{numerical range} $W_{\rm loc}(C,A)$ as a special
case of the relative $C$"~numerical range thus inheriting its mathematical
structure.

In particular detail, we analysed circular symmetry of local $C$"~numerical
ranges, which are of special interest in quantum control
and quantum information \cite{schuldihegla:07}. Here, we proved that local
$C$"~numerical ranges with circular symmetry have to be circular discs
centered at the origin of the complex plane. This is not evident as relative
$C$"~numerical ranges are in general not simply connected. Finally, we
applied our results to characterise all $(4 \!\times\! 4)$-matrices
with circular local $C$"~numerical range.

However, explicit formulas for the radius of a circular relative
$C$"~numerical range or, more general, for the 
{\em relative $C$"~numerical radius} of $A$
\begin{equation}\label{rCADef}%
r_K(C,A):=\max\big\{|\tr(C^\dagger UAU^\dagger)|\big|U \in K\}.
\end{equation}%
lead to open research problems. Therefore, general numerical algorithms for
finding sharp bounds on the size of $W_K(C,A)$ are highly desirable.
Geometric optimisation methods for the classical as well as the local
$C$"~numerical radius can be found in \cite{bro:88,helmke,glaser:98a,
dhkgsh:06,helmkehueper,dhksh:06,schuldihegla:07}.
Yet another interesting open problem is to find out conditions
which guarantee convergence of these methods, such as intrinsic
gradient flows, to the  relative $C$"~numerical radius
or at least to boundary points of the respective relative
$C$"~numerical range.
Problems of this kind are anticipated to be illuminating both
for mathematical structure and for quantum applications.


\section*{Acknowledgements}
This work has been supported in parts
by the German Research Foundation (DFG), grant HE 1858/10-1 KONNEW
as well as the integrated EU programme QAP.

\smallskip
\noindent
Parts of it were done while the third author was affiliated to 
National ICT Australia and the Department of Information Sciences and 
Engineering, The Australian National University, Canberra ACT 0200.

\smallskip
\noindent
National ICT Australia is funded by the Australian Government's 
Department of Communications,
Information Technology and the Arts and the Australian Research Council 
through {\em {Backing Australia's Ability}} and the ICT Research Centre 
of Excellence programs.




\appendices
\section{Two Technical Lemmas}
\label{appendix}

The following appendix contains two technical lemmas which we
referred to in the previous sections. Although most readers will be
familiar with these results, we incude them for completeness. 

\begin{lemma}
\label{lem:starshaped}
Let $W$ be a convex (or star-shaped) subset of a complex or
real vector space and let $[r,s]$ be a non-negative, real interval,
i.e. $0 \leq r \leq s$. Then the set
$[r,s] \cdot W := \{ \lambda \cdot w \;|\;  \lambda \in [r,s], w \in W\}$
is also convex (star-shaped).
\end{lemma}

\begin{proof}
For $s = 0$ there is nothing to prove thus we may assume without loss
of generality $s = 1$, as for $s > 0$ we can use the identity
$[r,s] \cdot W = [r/s,1] \cdot sW$ and the fact that $s W$ is
convex or star-shaped if $W$ is convex or star-shaped. 

Firstly, we consider the case that $W$ is convex. Let
$v_1, v_2 \in [r,1] \cdot W$, i.e. there are $w_1,w_2 \in W$
and $\lambda_1, \lambda_2 \in [r,1]$ such that $v_1 = \lambda_1 w_1$
and  $v_2 = \lambda_2 w_2$. We have to show
$$
v_1 + t (v_2 - v_1) \in [r,1] \cdot W
$$
for all $t \in [0,1]$. Without loss of generality let
$\lambda_1 \leq \lambda_2$ and define $\lambda^*$ and $t^*$ by
\begin{eqnarray*}
\lambda^* & := & \lambda_1 + t(\lambda_2 - \lambda_1) \\\\
t^* & := & 
\left\{
\begin{array}{lcl}
\frac{t \lambda_2}{\lambda_1 + t(\lambda_2 - \lambda_1)}
& \mathrm{for} & \lambda_1 \neq 0,\\\\
1 & \mathrm{for} & \lambda_1 = 0.
\end{array}
\right.
\end{eqnarray*}
As $t \in [0,1]$ it holds that $\lambda_1 \leq \lambda^* \leq \lambda_2$
and $0 \leq t^*  \leq 1$. Furthermore, we have
\begin{eqnarray}\label{eq2.2a}
\lambda^* \big(w_1 + t^*(w_2 - w_1)\big)
& = & 
\big(\lambda_1 + t(\lambda_2 - \lambda_1)\big)w_1 + t\lambda_2(w_2 - w_1)
\nonumber\\
& = & 
\lambda_1 w_1 + t(\lambda_2 w_2 - \lambda_1 w_1)\\
& = &
v_1 + t (v_2 - v_1).\nonumber
\end{eqnarray}
Hence $v_1 + t (v_2 - v_1)$ is in $[r,1] \cdot W$, as convexity of $W$
implies that the left side of Eq. (\ref{eq2.2a}) is obviously in
$[r,1] \cdot W$. 

\medskip
\noindent
Now, we assume that $W$ is star-shaped with star center $w_0$. Let
$v \in [r,1] \cdot W$, i.e. there is $w \in W$ and $\lambda \in [r,1]$
such that $v = \lambda w$. We have to show that there exists
a star center $v_0 \in [r,1] \cdot W$ such that
$$
v + t (v_0 - v) \in [r,1] \cdot W
$$
for all $t \in [0,1]$. Let $v_0 := w_0$ and define $\lambda^*$ and
$t^*$ in the same way as in Eq. (\ref{eq2.2a}) with
$\lambda_1 = \lambda$ and $\lambda_2 = 1$. Then we have
\begin{eqnarray}\label{eq2.2b}
\lambda^* \big(w + t^*(w_0 - w)\big)
& = & 
\lambda w + t(w_0 - \lambda w)
\; = \;
v + t (v_0 - v).
\end{eqnarray}
As $v_0$ does not depend on $v$, it follows that
$[r,1] \cdot W$ is star-shaped with star center $v_0$.
\end{proof}

\begin{lemma}
\label{lem:homotopy}
Let $\gamma: [a,b] \to \C$ be a closed curve and let
$\gamma^{-1}$ be its inverse, i.e. $\gamma^{-1}(t):=\gamma(b + a - t)$.
Moreover, let $z_0 \in \C$ be any point in the interior of $\gamma$,
i.e. the winding number of $\gamma$ with respect to $z_0$ is not equal
to zero. Then any homotopy from $\gamma$ to its inverse $\gamma^{-1}$
has to cross $z_0$.
\end{lemma}

\begin{proof}
Let $w(\cdot,z_0)$ denote the winding number of a closed curve
with respect to $z_0$ and assume that $h:[a,b] \times [c,d] \to \C$
is a homotopy from $\gamma$ to its inverse $\gamma^{-1}$ such
that $h(t,s) \neq z_0$ for all $(t,s) \in [a,b] \times [c,d]$.
As well-known, the winding number assumes only integer
values and satisfies the equality $w(\gamma,z_0) = - w(\gamma^{-1},z_0)$.
Therefore, by continuity of $h$, the winding number of
$\gamma$ with respect to $z_0$ has to be zero. This, in turn,
contradicts our hypothesis and thus $h$ has to cross $z_0$,
i.e. $h(t,s) = z_0$ for some $(t,s) \in [a,b] \times [c,d]$.
\end{proof}


\begin{thebibliography}{9}

\bibitem{ando:94}
T.~Ando and C.-K. Li, editors.
\newblock {\em Special Issue: The Numerical Range and Numerical Radius},
volume 37, 1--3 of {\em Linear and Multilinear Algebra}, pages 1--238.
\newblock Gordon and Breach, 1994.


\bibitem{bro:88}
R.W. Brockett.
\newblock Dynamical systems that sort lists, diagonalize matrices, and solve
  linear programming problems.
\newblock In {\em {Proc. {IEEE} of the 27th Conference on Decision and
  Control}}, 799--803, Austin, Texas, 1988.
\newblock {See also {\em Lin. Algebra \& Appl.}, 146:79-91, 1991}.

\bibitem{broedie}
T.~Br{\"o}cker and T. tom Dieck.
\newblock {\em Representations of Compact Lie Groups}.
\newblock Graduate Texts in Mathematics. Springer, New York, 1985.


\bibitem{cheung:96a}
W.-S. Cheung and N.-K. Tsing.
\newblock The {$C$}-numerical range of matrices is star-shaped.
\newblock {\em Linear and Multilinear Algebra}, 41:245--250, 1996.

\bibitem{daz:94}
J.~Dazord.
\newblock On the {$C$}-numerical range of a matrix.
\newblock {\em Lin. Algebra \& Applic.}, 212/213:21--29, 1994.

\bibitem{dhkgsh:06}
G. Dirr and  U. Helmke and M. Kleinsteuber and S. Glaser and
Th. Schulte--Herbr\"uggen.
\newblock The local numerical range: Examples, conjectures and
numerical algorithms
\newblock {\em Proc. MTNS}, Kyoto, 2006.

\bibitem{dhksh:06}
G. Dirr and  U. Helmke and M. Kleinsteuber and Th. Schulte--Herbr\"uggen.
\newblock A new type of {$C$}-numerical range arising in quantum computing
\newblock {\em PAMM, Proc. GAMM}, Berlin, 2006.

\bibitem{glaser:98a}
S.J. Glaser, T.~Schulte-Herbr{\"u}g\-gen, M.~Sieve\-king, O.~Schedletzky, N.C.
  Niel\-sen, O.W. S{\o}rensen, and C.~Griesinger.
\newblock Unitary control in quantum ensembles: Maximizing signal intensity in
  coherent spectroscopy.
\newblock {\em Science}, 280:421--424, 1998.

\bibitem{gusrao:97}
K.~Gustafson and D.~Rao.
\newblock {\em Numerical Ranges: The Field of Values of Linear Operators
and Matrices}. Universitext. \newblock Springer, New York, 1997.

\bibitem{gold:77a}
M.~Goldberg and E.G. Straus.
\newblock Elementary inclusion relations for generalized numerical ranges.
\newblock {\em Lin. Algebra \& Applic.}, 18:1--24, 1977.

\bibitem{haus:19}
D.~Hausdorff.
\newblock Der Wertevorrat einer Bilinearform.
\newblock {\em Math. Zs.}, 3:314--316, 1919.

\bibitem{helmke}
U.~Helmke and J.B. Moore.
\newblock {\em Optimization and Dynamical Systems}.
\newblock CCES. Springer, London, 1994.

\bibitem{helmkehueper}
U.~Helmke, K.~H{\"u}per, J.~B.~Moore, and Th.~Schulte-Herbr{\"u}ggen.
\newblock Gradient flows computing the {C-}numerical range with
applications in {NMR} spectroscopy.
\newblock {\em Journal of Global Optimization} 23: 283--308, 2002.

\bibitem{li:94}
C.-K. Li.
\newblock {$C$}-numerical ranges and {$C$}-numerical radii.
\newblock {\em Linear and Multilinear Algebra}, 37:51--82, 1994.

\bibitem{li:91}
C.-K. Li and N.-K. Tsing.
\newblock Matrices with circular symmetry on their unitary orbits and
  {$C$}-numerical ranges.
\newblock {\em Proceedings of the American Math. Soc.}, 111(1):19--28, 1991.

\bibitem{loja:84}
S.~{\L}ojasiewicz.
\newblock Sur les trajectoires du gradient d'une fonction analytique.
\newblock Seminari di geometria 1982-1983, {Universit{\`a} di Bologna,
Istituto di Geometria, Dipartimento di Matematica}, 1984.

\bibitem{vneumann}
J.~von Neumann.
\newblock {\em Mathematical Foundations of Quantum Mechanics}.
\newblock Princeton University Press, Princeton, 1955.

\bibitem{vneumann:37}
J.~von Neumann.
\newblock {Some Matrix-Inequalities and Metrization of Matrix-Space}.
\newblock {\em Tomsk Univ. Rev.}, 1:286--300, 1937.
\newblock [reproduced in: {\em John von Neumann: Collected Works},
A.H. Taub, Ed., Vol. IV: Continuous Geometry and Other Topics, 205-219,
Pergamon Press, Oxford, 1962.]

\bibitem{poon:80}
Y.~Poon.
\newblock Another proof of a result of Westwick.
\newblock {\em Linear and and Multilinear Algebra}, 9:35--37, 1980.

\bibitem{schulte:98}
T.~Schulte-Herbr{\"u}ggen.
\newblock {\em Aspects and Prospects of High-Resolution {NMR}}.
\newblock PhD thesis, ETH Z{\"u}rich, 1998.
\newblock Diss. ETH No. 12752.

\bibitem{schuldihegla:07}
T.~Schulte-Herbr{\"u}ggen, G.~Dirr, U.~Helmke and S.~Glaser
\newblock {\em The Significance of the $C$"~Numerical Range and the
Local $C$"~Numerical Range in Quantum Control and Quantum Information}.
\newblock accompanying paper in WONRA Proceedings.
\newblock e"~print: http://arxiv.org/pdf/math-ph/0701035, 2007.

\bibitem{toep:18}
R.~Toeplitz.
\newblock Das algebraische Analogon zu einem Satze von Fej{\'e}r.
\newblock {\em Math. Zs.}, 2:187--197, 1918.

\bibitem{west:75}
R.~Westwick.
\newblock A theorem on numerical range.
\newblock {\em Linear and Multilinear Algebra}, 2:311--315, 1975.


\end{thebibliography}
\end{document}